\documentclass[fleqn,usenatbib]{mnras}

\usepackage{newtxtext,newtxmath}
\usepackage{subcaption}
\usepackage[T1]{fontenc}

\DeclareRobustCommand{\VAN}[3]{#2}
\let\VANthebibliography\thebibliography
\def\thebibliography{\DeclareRobustCommand{\VAN}[3]{##3}\VANthebibliography}

\usepackage{graphicx}	
\usepackage{amsmath}	

\title[An `Almost' Dark Cloud]{WALLABY Pilot Survey: An `Almost' Dark Cloud near the Hydra Cluster}

\author[T. O'Beirne et al.]{
T. O'Beirne,$^{1,2}$\thanks{E-mail: tamsynobeirne.astro@gmail.com}
L. Staveley-Smith,$^{1,2}$
O. I. Wong$^{3,1,2}$
T. Westmeier,$^{1,2}$
G. Batten,$^{4}$
V. A. Kilborn,$^{4}$
\newauthor
K. Lee-Waddell,$^{1,3,5}$ 
P. E. Mancera Piña,$^{6}$
J. Román,$^{7,8}$
L. Verdes-Montenegro,$^{9}$
B. Catinella,$^{1,2}$ 
\newauthor
L. Cortese,$^{1,2}$
N. Deg,$^{10}$
H. Dénes,$^{11}$
B.~Q. For,$^{1,2}$
P. Kamphuis,$^{12}$
B. S. Koribalski,$^{13,14}$
C. Murugeshan,$^{3,2}$
\newauthor
J. Rhee,$^{1}$
K. Spekkens,$^{15}$
J. Wang,$^{16}$
K. Bekki,$^{1}$
and Á. R. López-Sánchez$^{17,18,2}$
\\
$^{1}$International Centre for Radio Astronomy Research (ICRAR), The University of Western Australia, 35 Stirling Highway, Crawley, WA 6009, Australia\\
$^{2}$ARC Centre of Excellence for All Sky Astrophysics in 3 Dimensions (ASTRO 3D), Australia\\
$^{3}$CSIRO Astronomy and Space Science, PO Box 1130, Bentley WA 6102, Australia\\
$^{4}$Centre for Astrophysics and Supercomputing, Swinburne University of Technology, Hawthorn, Victoria 3122, Australia\\
$^{5}$International Centre for Radio Astronomy Research (ICRAR), Curtin University, Bentley, WA 6102, Australia\\
$^{6}$Leiden Observatory, Leiden University, P.O. Box 9513, 2300 RA Leiden, The Netherlands\\
$^{7}$Kapteyn Astronomical Institute, University of Groningen, PO Box 800, 9700 AV Groningen, The Netherlands\\
$^{8}$Departamento de Astrof\'{\i}sica, Universidad de La Laguna, E-38206, La Laguna, Tenerife, Spain\\
$^{9}$Instituto de Astrofísica de Andalucía (CSIC)\\
$^{10}$Department of Physics, Engineering Physics, and Astronomy, Queen's University, Kingston, ON, K7L 3N6, Canada\\
$^{11}$School of Physical Sciences and Nanotechnology, Yachay Tech University, Hacienda San José S/N, 100119, Urcuquí, Ecuador\\
$^{12}$Ruhr University Bochum, Faculty of Physics and Astronomy, Astronomical Institute (AIRUB), 44780 Bochum, Germany\\
$^{13}$CSIRO Astronomy and Space Science, Australia Telescope National Facility, P.O. Box 76, NSW 1710, Australia\\
$^{14}$School of Science, Western Sydney University, Locked Bag 1797, Penrith, NSW 2751, Australia\\
$^{15}$Department of Physics and Space Science, Royal Military College of Canada, PO Box 17000, Station Forces, Kingston, Ontario, Canada, K7K 7B4\\
$^{16}$Kavli Institute for Astronomy and Astrophysics, Peking University, Beijing, 100871, China\\
$^{17}$School of Mathematical and Physical Sciences, Macquarie University, NSW 2109, Australia\\
$^{18}$Macquarie University Research Centre for Astrophysics and Space Technologies, NSW 2109, Australia\\ 
}

\date{Accepted XXX. Received YYY; in original form ZZZ}

\pubyear{2022}

\begin{document}
\label{firstpage}
\pagerange{\pageref{firstpage}--\pageref{lastpage}}
\maketitle

\begin{abstract}
We explore the properties of an `almost' dark cloud of neutral hydrogen (H{\sc i}) using data from the Widefield ASKAP L-band Legacy All-sky Survey (WALLABY). Until recently, WALLABY J103508-283427 (also known as H1032-2819 or LEDA 2793457) was not known to have an optical counterpart, but we have identified an extremely faint optical counterpart in the DESI Legacy Imaging Survey Data Release 10. We measured the mean $g$-band surface brightness to be $27.0\pm0.3$ mag arcsec$^{-2}$. The WALLABY data revealed the cloud to be closely associated with the interacting group Klemola~13 (also known as HIPASS J1034-28 and the Tol~9 group), which itself is associated with the Hydra cluster. In addition to WALLABY J103508-283427/H1032-2819, Klemola~13 contains ten known significant galaxies and almost half of the total H{\sc i} gas is beyond the optical limits of the galaxies. By combining the new WALLABY data with archival data from the Australia Telescope Compact Array (ATCA), we investigate the H{\sc i} distribution and kinematics of the system. We discuss the relative role of tidal interactions and ram pressure stripping in the formation of the cloud and the evolution of the system. The ease of detection of this cloud and intragroup gas is due to the sensitivity, resolution and wide field of view of WALLABY, and showcases the potential of the full WALLABY survey to detect many more examples.

\end{abstract}




\section{Introduction}

With the new generation of deep, large area surveys, e.g. the Arecibo Legacy Fast ALFA \citep[ALFALFA,][]{Giovanelli:05}, the DESI Legacy Imaging Survey Data Release 10 \citep[hereafter refered to as the Legacy Survey DR10;][]{Dey:19}, the low surface brightness universe is beginning to be explored. Ultra-diffuse galaxies (UDGs) are extremely faint and diffuse galaxies. There are several proposed definitions \citep[e.g.,][]{Koda:15,Lim:20}, with one of the most commonly used definitions being that of \cite{VanDokkum:15}: galaxies with a central surface brightness $\mu_{0,g} \ge 24$ mag arcsec$^{-2}$ and an effective radius $r_{eff} \ge 1.5$ kpc. \cite{VanDokkum:15} proposed that UDGs are `failed' galaxies that were prevented from becoming typical galaxies due to a loss of their gas supply early in their lifetime. In recent years, UDGs have been observed to have a range of properties and environments, from the red cluster UDGs of \cite{VanDokkum:15,Koda:15,vanderburg:16} and \cite{mancerapina:19} to the isolated H{\sc i} bearing UDGs (HUDs) of \cite{Leisman:17} and \cite{ManceraPina:20}. In order to explain the variety in observed properties, various formation mechanisms have been theorised including internal mechanisms, mergers and tidal interactions \citep[see][for summaries]{Jiang:19,Sales:20,Jones:21,LaMarca:22}. \cite{LaMarca:22} found 11 new UDGs within 0.4$r_{vir}$ of the Hydra Cluster, bringing the total number of UDGs in the Hydra cluster to 21, including the UDGs found in \citet{Iodice:20} and \citet{Iodice:21}. Due to their diffuse nature, UDGs have historically been very difficult to detect, but with recent surveys this is changing. Using the Legacy Survey, \cite{Zaritsky:22} found 5598 UDG candidates which make up the Systematically Measuring Ultra-Diffuse Galaxies (SMUDGes) catalogue.


At the extreme end of low surface brightness galaxies, lie dark galaxies. Dark galaxies are isolated, dark matter dominated systems without optical counterparts. They are predicted by $\Lambda$ cold dark matter cosmology \citep[e.g.,][]{BoylanKolchin:11}, and those that contain H{\sc i} have the potential to be detected by radio telescopes. However, they have been shown to be rare by recent studies \citep[e.g.,][]{Kilborn:00,Janowiecki:15,Bilek:20,Wong:21}, and their existence is even contested \citep[e.g.,][]{Bekki:05,Taylor:05}. One of the first arguments for the existence of `almost' dark galaxies came from \cite{Disney:76}. He argued that extremely low surface brightness galaxies were abundant in the universe but remained below the detection limits of the instruments available at the time. He postulated that low mass dwarf elliptical galaxies could just be the tip of giant low surface brightness spirals, dubbing these objects "crouching giants" \citep{Disney:87}. Modern cosmological simulations support the existence of dark galaxies and extremely low surface brightness galaxies. The presence of dark matter halos without significant stellar mass (dark galaxies) is a common explanation of the `missing satellites' and `too big to fail' problems of cosmological dark matter only simulations \citep{Klypin:99,BoylanKolchin:11,Sawala:16}. 

Although both are devoid of stars, it is useful to distinguish dark galaxies, which contain dark matter, from dark H{\sc i} clouds, which predominantly contain gas. Cosmological simulations have shown that dark galaxies can have a primordial origin \citep[e.g.,][]{Verde:02,Taylor:16}. In this scenario they are failed galaxies - dark matter halos that accumulate gas that never condenses and undergoes star formation, and are often isolated. HI1225+01 in the Virgo Cluster contains a promising example of a dark galaxy candidate, with a surface brightness limit of $\mu_{r,AB} > 28$ mag arcsec$^{-2}$ \citep{Matsuoka:12}. On the other hand, gas-dominated clouds can form as the result of tidal stripping, such as in the Leo ring \citep{Schneider:83}, and the dark clouds discussed in \cite{kilborn:06}, \cite{Cannon:15} and \cite{Taylor:22}. \cite{LeeWaddell:14} present their Giant Meter Wave Radio Telescope (GMRT) observations of a gas-rich interacting group that contains a H{\sc i} cloud with tidal origin. No optical counterpart had been detected in prior studies, but faint optical features were identified in their observations with the Canada-France-Hawaii Telescope. Clouds can also form due to ram pressure stripping, such as the plume of gas stripped from NGC 4388 in the Virgo cluster \citep{Oosterloo:05}. Both tidal interactions and ram pressure stripping could have played a role in the origin of the potentially old tidal dwarf galaxy studied in \cite{Duc:07}. Often it is not possible to definitively determine whether a H{\sc i} source has a primordial or interaction based origin, such as the dark clouds detected in the Eridanus group  by \cite{Wong:21}. VIRGOHI21 was initially identified in the H{\sc i} Jodrell All-Sky Survey \citep[HIJASS][]{Lang:03} data as a dark galaxy candidate \citep{Davies:04,Minchin:05}. Additional data from Arecibo led \cite{Haynes:07} to conclude that the dark source actually formed via harassment of NGC 4254. Subsequently, the system was modelled by \cite{Duc:08}, who show that instead of a primordial origin, the source may have been formed by a high velocity encounter between two galaxies. 

Tidal dwarf galaxies (TDG) form from the debris left by mergers and interacting galaxies. They are self-gravitating and do not contain a significant amount of dark matter \citep[e.g.,][]{Bournaud:06}. In their study of TDGs in the local Universe, \cite{Kaviraj:12} found that TDGs have stellar masses less than $10\%$ of their parent galaxies and lie within 15 optical half-light radii of the parent galaxies. They also found that they contain both old stars drawn from their parent galaxies, as well as newly formed stars. \cite{Gray:23} analysed several `almost' dark galaxies from ALFALFA and found two TDGs at a later stage of their evolution, lying further from their potential parent galaxies. The ALFALFA almost dark galaxy studied in \cite{Leisman:21} is also shown to likely be a TDG. \cite{Roman:21} show that a TDG in Hickson Compact Group 16 has similar properties to UDGs and with $\sim2$ Gyr of evolution could become undetectable in the optical, matching the observational properties of a dark galaxy.

H{\sc i} observations are particularly useful for investigating the evolution of galaxies as H{\sc i} is typically more extended than the optical disc. H{\sc i} is also a good tracer of both tidal interactions and ram pressure stripping. \cite{Taylor:05} found that in all their models dark galaxies would have detectable H{\sc i} emission. However, due to poor angular resolution and source confusion, it was difficult to definitively identify dark galaxies in the H{\sc i} Parkes All Sky Survey \citep[HIPASS,][]{Barnes:01, Doyle:05}. Less then $2\%$ of extragalactic ALFALFA sources lack an optical counterpart \citep{Haynes:18}, and many of these dark objects are associated with tidal debris \citep[e.g.,][]{Leisman:16}, with a few dark galaxy candidates followed up \citep[e.g.,][]{Kent:10}.  With the improved resolution and sensitivity of the Australian Square Kilometre Array Pathfinder \citep[ASKAP,][]{Hotan:21} the Widefield ASKAP L-band Legacy All-sky Survey \citep[WALLABY,][]{Koribalski:20} has the potential to detect dark galaxy candidates and extremely low surface brightness galaxies by being able to better locate the origin of the emission and better separate emission from other nearby objects in denser group and cluster environments. The fast survey speed will also allow more of these rare objects to be detected in the large volume covered. Inspired by the dark clouds detected in the pre-pilot data from WALLABY by \cite{Wong:21}, this study began as a search for dark galaxy candidates in the phase 1 pilot WALLABY data \citep{Westmeier:22}, with six candidates being found in the Hydra field alone. Being mostly close to the noise level, follow-up observations of the candidates are necessary to confirm their reality, or otherwise. However, one of the candidates, WALLABY J103508-283427, has previously been detected by the Karl Jansky Very Large Array (VLA), the Nançay telescope,  and the Australia Telescope Compact Array (ATCA), and is therefore a genuine detection. This paper presents analysis of  WALLABY and archival data for this dark galaxy candidate alone. 


WALLABY J103508-283427 was first detected with the VLA by \cite{McMahon:93} and subsequently with Nançay  \citep{Duc:99}, as H1032-2819.  WALLABY \citep{Westmeier:22} and archival ATCA data \citep{LopezSanchez:08} reveal that this H{\sc i} cloud is part of a complex interacting system in the Klemola~13 galaxy group \citep{Klemola:69} (also known as HIPASS J1034–28 and the Tol 9 group). Klemola~13 is thought to contain the following galaxies: ESO~436-G046, ESO~436-IG042, ESO-LV~4360421, ESO~436-G044, ESO~436-G045, PGC~031288, PGC~031270 and ESO 437-G001 based on their similar redshifts and small projected distances. Additionally, ESO~437-G004 and 2MASX J10355262-2817269 are considered to be more distant members of the group.  A detailed WALLABY study of the intragroup gas and other potential clouds in the whole system is underway by Batten et al. (in preparation). \cite{LopezSanchez:08} found using their ATCA data that the H{\sc i} gas in this group is predominantly located in two regions: one around the spiral galaxy ESO~436-G046, and one around ESO~436-IG042 (Tol 9), with the maximum H{\sc i} column density found in ESO~436-IG042. They also noted an extended arm to the north in the direction of ESO~436-G044 and ESO~436-G045, but did not detect any emission from these galaxies. No previous studies have been able to identify an optical counterpart for WALLABY J103508-283427/H1032-2819 (hereafter H1032-2819). 

In this paper we study the properties and environment of this system in depth with the addition of the new WALLABY data, and identify for the first time an extremely faint optical counterpart using the Legacy Survey DR10. The structure of this paper is as follows. In Section \ref{sec:obs}, we describe the WALLABY and archival observations, followed by Section \ref{sec:process} which outlines the data reduction and source finding process. The main observational results of this study are presented in Section \ref{sec:results}, including the H{\sc i} distribution and kinematics of the system, and its optical properties. Section \ref{sec:discuss} discusses the evolution of the almost dark H{\sc i} cloud and the group with which it appears to be associated. Section \ref{sec:concl} summarises our findings. For the purposes of this paper, the distance to the Hydra cluster is 47.5 Mpc \citep{Kourkchi:17} and Klemola~13 is set to be at the same distance. The Tully-Fisher distances for the galaxies in Klemola~13 are clustered around a similar value. Throughout the paper we adopt the AB magnitude convention.


\section{Data}
\label{sec:obs}
\subsection{ASKAP Observations}

This study uses data from the Hydra field of phase 1 of the WALLABY pilot survey \citep{Westmeier:22}. WALLABY is a H{\sc i} survey of the local Universe ($z<0.1$) which has been allocated 8832 hours of observing time over the next five years. WALLABY is being conducted with ASKAP \citep{Hotan:21}, a radio interferometer composed of 36 $\times$ 12-m dishes equipped with Phased-Array Feeds and located in Western Australia. For reasons of surface brightness sensitivity and computing requirements, the main WALLABY ASKAPsoft pipeline does not use baselines over 2 km, although a high-resolution cutout mode for pre-defined targets is currently being implemented (Murugeshan et al., in preparation). With the baselines longer than 2 km removed, the angular resolution is 30\arcsec.


\subsection{Archival ATCA Data}

Archival data from ATCA \citep[see][]{LopezSanchez:08} were also used in this study. ATCA is an Earth rotation aperture synthesis radio interferometer, which consists of 6 $\times$ 22-m dishes, located in New South Wales, Australia.  Table \ref{tab:ATCA} shows the start date and integration time for each of the 5 array configurations used in the observations. PKS B1934-638 was used as the primary calibrator and PKS B1015-314 was used as the secondary calibrator. Across all configurations, a total of 50.2 hours was spent on source.  The centre frequency of the observations was 1.405~GHz with a bandwidth of 8 MHz and a frequency resolution of 15.6~kHz. The ATCA primary beam full width at half maximum (field of view) is 34.5\arcmin. The observations were carried out in 2007 and 2008 with the pre-CABB correlator.

        \label{fig:uvatca}

\begin{table}
    \centering
    \caption{Observation details of the archival ATCA data \citep{LopezSanchez:08}. The time given is the on source integration time.}
    \label{tab:ATCA}
    \begin{tabular}{lccc}
    \hline
        Array configuration & Date & Time & Total Time\\
         &  & (h) & (h)\\
        \hline
        EW352 & 2008 Feb 10 & 7.0 & 7.0 \\
        & \\
         EW367 & 2008 Nov 20 & 5.3 & 7.3 \\
         & 2008 Nov 21 & 2.0 &  \\
        & \\

        750A & 2007 Jan 28 & 9.7 & 9.7\\
        & \\

        1.5A & 2007 Jan 29 & 7.6 & 16.6\\
         & 2007 Jan 30 & 9.0 & \\
        & \\

        6A & 2008 Nov 06 & 9.6 & 9.6 \\
        & \\

        Total &&& 50.2 \\
        \hline
    \end{tabular}
    
\end{table}

\subsection{Deep Optical Images}

We use data from the DESI Legacy Imaging Surveys \citep{Dey:19}. This survey is mainly carried out with the 4 m Blanco Telescope, providing imaging in $g, r, i$ and $z$ bands over a region of $> 20 000$ square degrees with seeing of the order of 1 arcsec. We use the most recent Data Release 10. The images were obtained from the public website of the survey (https://www.legacysurvey.org/). The limiting surface brightnesses were calculated on the images, with values: 29.8, 29.4, 27.7 and 28.0 mag arcsec$^{-2}$ in the  $g, r, i$ and $z$ bands respectively, measured in $3\sigma$, $10\times10$ arcsec boxes following the depth definition by \cite{ROman:20}. We note the considerably higher depth of the $g$ and $r$ bands compared to the $i$ and $z$ bands. These ancillary data sets are discussed in Sect. \ref{sec:results}.

\section{Analysis}
\label{sec:process}
\subsection{ASKAP}

The data were reduced and calibrated using the ASKAPsoft pipeline \citep{Cornwell:11,Guzman:19,Wieringa:20,Whiting:20}.  The spectral image cubes for each primary beam were deconvolved using the multi-scale CLEAN algorithm \citep{Cornwell:08,Rau:11}. See \cite{Westmeier:22} for a more detailed description of the data reduction and calibration process. The rms in the final datacube is 1.6~mJy~beam$^{-1}$, corresponding to a $3\sigma$ column density sensitivity of $1.18\times10^{20}$~cm$^{-2}$ over 20~km~s$^{-1}$. We apply the statistical correction to the integrated fluxes that we measure to account for the flux underestimation as outlined in \cite{Westmeier:22}.

\subsection{ATCA}

We used the {\tt MIRIAD} software package \citep{Sault:95} to process the archival ATCA data. Each configuration was reduced separately and combined using the {\tt INVERT} task. To reduce the data, first we flagged any radio frequency interference using {\tt UVFLAG}, {\tt BLFLAG} and {\tt PGFLAG}. Once we were satisfied with the quality of the data, the bandpass and phase calibration were conducted with the primary calibrator PKS B1934-638 and the secondary calibrator PKS B1015-314 using the tasks {\tt MFCAL}, {\tt MFBOOT}, {\tt GPCAL} and {\tt GPBOOT}. Then the continuum subtraction was performed with {\tt UVLIN}. After this, the visibility data for each baseline was combined using the {\tt INVERT} task, which Fourier transforms the UV data to create the dirty image. We used a cell size of 5\arcsec, an image size of 512 by 512 pixels, a velocity channel width of 20 km s$^{-1}$ and a robust parameter of 0.5. Antenna 6 was removed to allow us to recover the low column density intragroup gas. Longer baselines enable higher angular resolution imaging at the cost of surface brightness sensitivity.  As Antenna 6 is an outlier, the resulting long baselines may not be desirable for imaging diffuse low surface brightness emission. Hence, the removal of antenna 6 will help us recover the emission on the scales that are relevant for this paper. The velocity is in the optical convention ($cz$) and in the heliocentric frame. For the data available and the weighting used, the estimated theoretical rms noise is 0.82 mJy. Next the deconvolution was performed using the {\tt CLEAN} task, using the default {\tt MIRIAD} hybrid Högbom/Clark/Steer algorithm with a cutoff of 7 mJy. The final image cube was created from the dirty image and the dirty beam using {\tt RESTOR} and the primary beam correction was applied using {\tt LINMOS}. The rms level of the final processed data cube at the pointing centre is 0.75 mJy, corresponding to a $3\sigma$ column density sensitivity of $1.44\times10^{19}$ cm$^{-2}$ over 20 km s$^{-1}$. As the noise is not constant across the field, the rms at the centre of each aperture can be found in Table \ref{tab:results}. The synthesised beam is 1\farcm52 $\times$ 0\farcm63 at a position angle of $-2\fdg8$.

\subsection{Source Finding}

Version 2 of the Source Finding Application (SoFiA 2) \citep{Serra:15,Westmeier2:21} was used for source finding for both the ASKAP and ATCA data. In the automated pipeline used in the pilot phase 1 data release of WALLABY \citep{Westmeier:21}, this complex interacting system was split into 3 separate detections and excluded some of the fainter intragroup emission. Consequently, we reran SoFiA 2 with parameters that allowed us to recover more extended emission than the automated pipeline. Table \ref{tab:SoFiA} outlines the important SoFiA parameter settings in the S+C (smooth and clip) finder, linker and reliability modules for the ATCA data, the automated WALLABY pipeline and our re-run. For details on these parameters see \cite{Serra:15} and \cite{Westmeier2:21}. Batten et al. (in preparation) use different SoFiA parameters in order to better identify cloud candidates within the extended group emission. In all further analysis, unless otherwise stated, the WALLABY data with the SoFiA mask generated for this work has been used. 

\begin{table}
    \centering
    \caption{SoFiA parameter settings for the S+C finder ({\tt SCFIND}), linker and reliability modules.}
    \label{tab:SoFiA}
    \setlength{\tabcolsep}{2.5pt}
    \begin{tabular}{lccc}
    \hline
        Parameter & WALLABY & ATCA   & WALLABY  \\
         & \cite{Westmeier:22} & this paper & this paper \\
        \hline
        scfind.kernelsXY & 0, 5, 10 & 0, 5, 10  & 0, 5, 10, 15 \\
        scfind.kernelsZ & 0, 3, 7, 15 & 0, 3, 7, 15 & 0, 3, 7, 15, 31 \\
        scfind.threshold & 3.5& 3.9  & 3.5 \\
        linker.radiusXY & 2 & 2 & 2 \\
        linker.radiusZ & 2 & 2 & 2 \\
        linker.minSizeXY & 8 & 8 & 0 \\
        linker.minSizeZ & 5 & 5 & 0 \\
        reliability.threshold & 0.8 & 0.7  & 0.9 \\
        reliability.scaleKernel & 0.4 & 0.4 & 0.4 \\
    \hline
    \end{tabular}
    
\end{table}


\section{Results}
\label{sec:results}
\subsection{H{\sc i} Properties and Distribution}
\label{sec:HIprop}

\begin{figure*}
    \centering
    \includegraphics[width=0.9\textwidth]{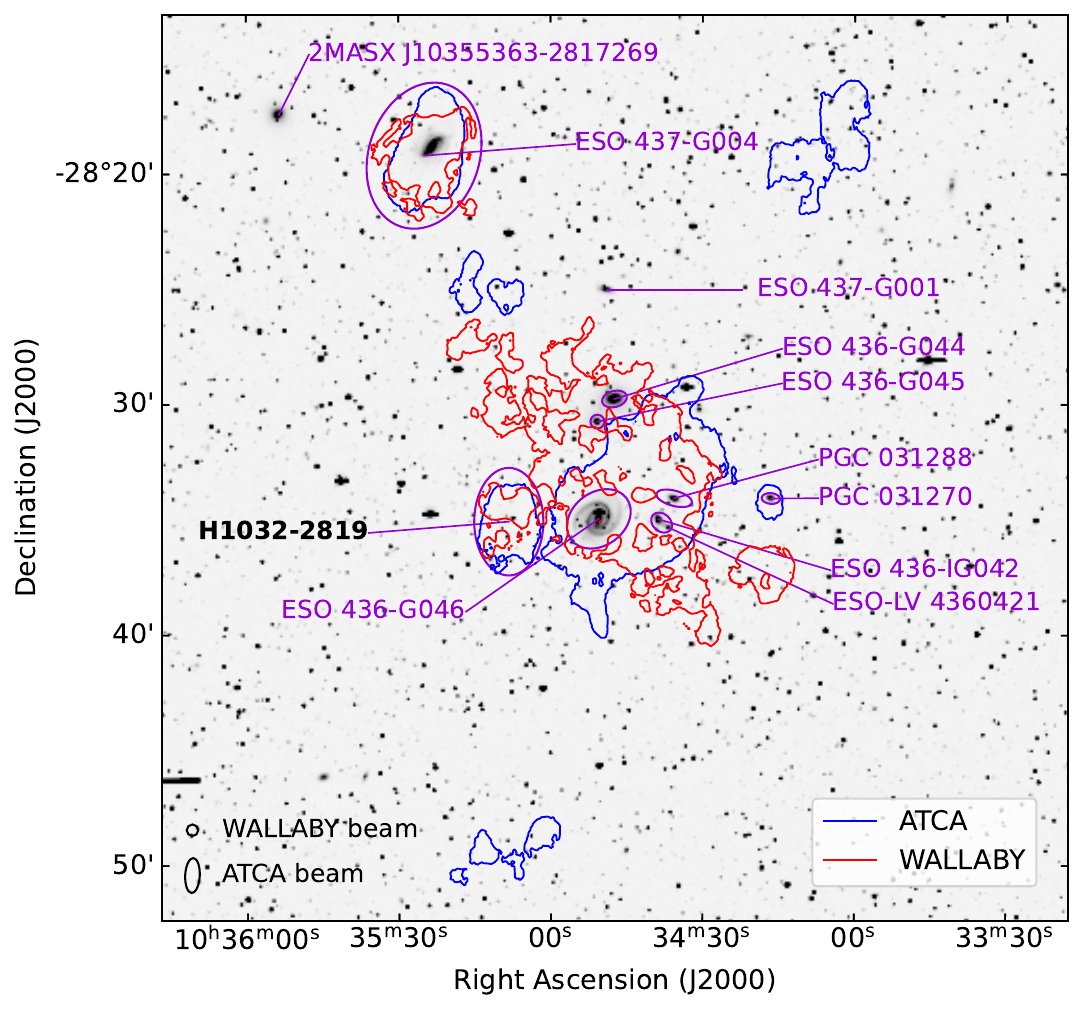}
    \caption{The location of the Klemola~13 galaxies and their optical apertures as defined in Section \ref{sec:HIprop} (in purple) are shown on the $g$-band Legacy Survey DR10 image.  Superposed on this image are the $1\times10^{19}$cm$^{-2}$ H{\sc i} column density contour level, representing the boundary of the SoFiA masks.  Coordinates, major and minor axes, and position angles are given in Table \ref{tab:results}. The tip of the purple lines points at the centre coordinates from Table \ref{tab:results}, except for those of ESO~436-IG042 and ESO-LV~4360421 which point the the respective galaxies. The almost dark cloud (H1032-2819; bold, black font), is detected in both the ATCA and WALLABY data. The ATCA and WALLABY beams are shown at the bottom left.}
    \label{fig:labels}
\end{figure*}

The overall column density map of Klemola~13 created with the SoFiA mask is shown in Figure~\ref{fig:labels}, superposed on a Legacy Survey DR10 $g$-band image. The contours cover eight galaxies at similar redshifts (with velocities within 405 km s$^{-1}$ of each other): ESO~436-G046, ESO~436-IG042, ESO-LV~4360421, PGC~031288, PGC~031270, ESO~436-G044, ESO~436-G045 and ESO~437-G004, as well as the almost dark cloud, H1032-2819. ESO~437-G001 and 2MASX J10355363-2817269 are at the same redshift as the rest of the system, but no H{\sc i} flux is contained in the masks used here for the WALLABY and ATCA data. Figure~\ref{fig:labels} also contains two H{\sc i} clouds detected by ATCA without optical counterparts at the north western and southern edges of the image. Whilst we can neither confirm nor deny whether these clouds are genuine, it is unlikely that they are real as they lie at the edge of the ATCA field where the noise increases significantly and are undetected by WALLABY. Hence, although they are shown here for completeness, they have not been included in any subsequent analysis. 

Coordinates, optical velocities, major and minor axes, and position angles for all sources with detected H{\sc i} in Klemola~13 can be found in the top section of Table~\ref{tab:results}. As H1032-2819 and ESO~437-G004 have a distinct separation (spatially and in velocity), elliptical apertures were created that enclosed the H{\sc i} gas contained in the SoFiA masks for both the ATCA and WALLABY data (see Figure \ref{fig:masks} for the masks). The position angles shown in Table~\ref{tab:results} are the position angles of these apertures. The rest of the galaxies are surrounded by shared intragroup gas that can't be distinctly associated with any one galaxy, so the elliptical apertures are obtained from their optical/IR dimensions. ESO 436-IG042 and ESO-LV~4360421 are undergoing a significant interaction that is clearly visible in the optical \citep{LopezSanchez:08}. Consequently, an elliptical aperture was created that encloses the optical emission of both of these galaxies, using the position angle from \cite{LopezSanchez:08}. For these galaxies, coordinates, major and minor axes were derived from the apertures. The coordinates of the rest of the sources are taken from the Two Micron All Sky Survey Extended objects (2MASX) catalogue \citep{2MASX:03} and the optical velocities are from the 6dF Galaxy Survey (6dFGS) Data Release 3 \citep{Jones:09}.  The major and minor axes, and the position angle of the apertures for the other four ESO galaxies (ESO~436-G046, ESO~436-IG042, ESO~436-G044 and ESO~436-G045) are taken from \cite{ESOups:82}. As these values were not available from \cite{ESOups:82} for PGC~031288 and PGC~031270, the major and minor axes, and position angle were derived from the values in the 2MASX catalogue. For all the other sources, the ESO optical values were on average $\sim5$ times larger than their respective $K$-band 2MASX values. Hence, a factor of 5 was used to convert the $K$-band major and minor axes into an ESO-equivalent optical value for these two galaxies.


Using these optical apertures, the flux, central velocity, velocity width and H{\sc i} mass were calculated for each galaxy (and the almost dark cloud) in the system in Table \ref{tab:results}. The central velocities measured from the ATCA data are systematically lower than those measured from the WALLABY data, particularly for the large face on spiral galaxy ESO~436-G046.  The different beam sizes and sensitivities also result in some of the smaller galaxies only being detected by one instrument. ESO~436-G044 and ESO~436-G045 were not detected by ATCA and PGC~031270 was not detected by WALLABY, so $3\sigma$ upper limits to the integrated flux were calculated. The two galaxies that contain the most H{\sc i} are ESO~436-G046 and ESO~437-G004, the latter has a central velocity most similar to H1032-2819. The total flux and H{\sc i} masses for the entire system are shown in Table \ref{tab:groupgas}, where the `intragroup gas' refers to the H{\sc i} gas in the entire system that is not contained within the apertures.  A significant proportion of the gas in the system is between the galaxies rather than within their optical apertures, with only $59\% / 52\%$ (WALLABY/ATCA) of the H{\sc i} contained within the apertures, with the limitation that optically defined apertures have been used for most of the galaxies. 


\begin{table*}
    \centering
    \caption{Key parameters for the interacting system Klemola~13 and its different components. From top to bottom the rows are: right ascension, declination, optical velocity, position angle, major diameter, minor diameter, stellar mass, specific star formation rate, integrated flux, H{\sc i} Mass, H{\sc i} velocity ($cz$), velocity width, rotational velocity, dynamical mass, H{\sc i} to stellar mass ratio, H{\sc i} to dynamical mass ratio, stellar to dynamical mass ratio, and the rms centred at the source (only ATCA). }
    \label{tab:results}
    \setlength{\tabcolsep}{4pt}
    \begin{tabular}{clcccccccc}
    \hline
         & & H1032-2819 & ESO  & ESO &  PGC & PGC & ESO  & ESO & ESO   \\
        & &   & 436-G046 & 436-IG042 &  031288 & 031270 & 436-G044 & 436-G045 & 437-G004   \\
        &&&& + LV4360421 &&&&& \\
        \hline
        & RA (J2000)  & 158\fdg785 & 158\fdg711 & 158\fdg659 & 158\fdg648 & $158\fdg569$ & 158\fdg700 & 158\fdg710 & 158\fdg854  \\
        & Dec (J2000) & $-28\fdg585$ & $-28\fdg583$ & $-28\fdg586$  & $-28\fdg568$ & $-28\fdg568$ & $-28\fdg500$ & $-28\fdg510$ & $-28\fdg320$  \\
        & $v_{\rm opt}$ (km s$^{-1}$) & - & $3390\pm45$ & $3445\pm45$ &  $3542\pm45$ & $3570\pm45$ & $3165\pm17$ &$3402\pm17$& $3318\pm45$\\
        & PA & $0^{\circ}$ & $127^{\circ}$ & $49^{\circ}$  & $75^{\circ}$ & $85^{\circ}$ & $105^{\circ}$ & $0^{\circ}$ & $162^{\circ}$ \\
        & $D_{\rm maj}$  &  280\arcsec & 180\arcsec & 96\arcsec  & 76\arcsec & 48\arcsec & 66\arcsec & 36\arcsec & 388\arcsec \\
        & $D_{\rm min}$  & 180\arcsec & 138\arcsec & 46\arcsec  & 40\arcsec & 28\arcsec & 42\arcsec & 36\arcsec & 288\arcsec \\
        \hline
        & $\log(M_{*}/$M$_{\odot})$ & $8.0\pm0.7$ & $9.8\pm0.1$ & $9.4\pm0.1$ & $9.8\pm0.1$ & $9.1\pm0.1$ & $10.2\pm0.1$ & $9.9\pm0.1$ & $9.7\pm0.1$ \\
        & $\log$(sSFR/yr$^{-1}$) & $<-9.6$ & $-9.6\pm0.2$ & $-9.1\pm0.2$ & $-12.7\pm0.2$ & $-10.0\pm0.2$ & $-11.9\pm0.2$ & $-12.0\pm0.2$ & $-10.0\pm0.2$ \\
        \hline
         & $S$ (Jy km s$^{-1}$) & $1.37\pm0.14$ & $3.72\pm0.21$ & $1.99\pm0.06$ & $1.20\pm0.09$ & $<0.29$ & $0.71\pm0.08$ & $0.41\pm0.05$ & $7.23\pm0.29$  \\
         & $\log(M_{\rm HI}/$M$_{\odot})$ & $8.86\pm0.04$ & $9.3\pm0.02$ & $9.03\pm0.01$ & $8.81\pm0.03$ & - & $8.58\pm0.05$ & $8.34\pm0.05$ & $9.59\pm0.02$\\
          & $v_{\rm c}$ (km s$^{-1}$) & 3286$\pm10$ & 3515$\pm20$ & 3473$\pm10$ & 3441$\pm10$ & - & 3555$\pm20$ & - & 3276$\pm10$ \\
         & $w_{50}$ (km s$^{-1}$) &  47$\pm20$ & 226$\pm20$ & 237$\pm20$ & 183$\pm20$ & - & 216$\pm40$ & - &307$\pm20$ \\  
         WALLABY & $v_{\rm rot}$ (km s$^{-1}$) & - & $155\pm30$ & - & - & - & - & - & $118\pm20$  \\ 
          & $\log(M_{\rm Dyn}/$M$_{\odot})$ & $10.1\pm0.4$ & $11.0\pm0.1$ & $11.0\pm0.1$ & $10.6\pm0.1$ & - & $10.8\pm0.1$ & - & $10.8\pm0.1$ \\ 
         & $\log(M_{\rm HI}/M_{*})$ & $0.9\pm0.3$ & $-0.5\pm0.1$ & $-0.5\pm0.1$ & $-1.0\pm0.1$ & - & $-1.6\pm0.1$ & $-1.5\pm0.1$ & $-0.1\pm0.1$\\ 
         & $\log(M_{\rm HI}/M_{\rm Dyn})$ & $-1.2\pm0.4$ & $-1.7\pm0.1$ & $-2.0\pm0.1$ & $-1.8\pm0.1$ & -& $-2.2\pm0.1$ & - & $-1.2\pm0.1$  \\ 
         & $\log(M_{*}/M_{\rm Dyn})$ & $-2.1\pm1.7$ & $-1.2\pm0.2$ & $-1.4\pm0.2$ & $-0.7\pm0.2$ & - & $-0.6\pm0.2$ & - & $-1.1\pm0.2$  \\ 
         \hline
         &  $S$ (Jy km s$^{-1}$) &  $0.92\pm0.13$ & $4.32\pm0.18$ & $1.01\pm0.04$ & $1.16\pm0.06$ & $0.06\pm0.02$ & $<0.27$ & $<0.18$ & $6.72\pm0.45$  \\
         & $\log(M_{\rm HI}/$M$_{\odot})$ &  $8.69\pm0.06$ & $9.36\pm0.02$ & $8.73\pm0.02$ & $8.79\pm0.02$ & $7.52\pm0.13$ & - & - & $9.55\pm0.03$  \\
          & $v_{\rm c}$ (km s$^{-1}$) &  3266$\pm10$ & 3421$\pm10$ & 3448$\pm10$ & 3429$\pm10$ & 3540$\pm10$ & - & - & 3237$\pm10$ \\
         & $w_{50}$ (km s$^{-1}$) &  57$\pm20$ & 238$\pm20$ & 227$\pm20$ & 179$\pm20$ & 109$\pm20$ & - & - & 291$\pm20$ \\
        ATCA & $v_{\rm rot}$ (km s$^{-1}$) & - & $142\pm30$ & - & - & - & - & - & $116\pm20$\\ 
         & $\log(M_{\rm Dyn}/$M$_{\odot})$ & $10.2\pm0.3$ & $11.0\pm0.2$ & $10.9\pm0.1$ & $10.5\pm0.1$ & $10.0\pm0.1$ & - & - & $10.8\pm0.1$ \\ 
         & $\log(M_{\rm HI}/M_{*})$ & $0.7\pm0.3$ & $-0.4\pm0.1$ & $-0.8\pm0.1$ & $-1.0\pm0.1$ & $-1.6\pm0.2$ & - & - & $-0.1\pm0.1$   \\ 
         & $\log(M_{\rm HI}/M_{\rm Dyn})$ & $-1.5\pm0.3$ & $-1.6\pm0.2$ & $-2.2\pm0.1$ & $-1.8\pm0.1$ & $-2.4\pm0.3$ & - & - & $-1.2\pm0.1$ \\ 
         & $\log(M_{*}/M_{\rm Dyn})$ & $-2.3\pm1.1$ & $-1.2\pm0.2$ & $-1.4\pm0.2$ & $-0.7\pm0.2$ & $-0.8\pm0.2$ & - & - & $-1.1\pm0.2$  \\ 
         & rms (mJy beam$^{-1}$) & 0.85 & 0.80 & 0.76 & 0.77 & 0.75 & 0.82 & 0.80 & 1.95 \\
         \hline
    \end{tabular}
        
\end{table*}

\begin{table}
    \centering
    \caption{Flux and H{\sc i} mass of the entire interacting system compared to that of the intragroup gas outside of the galaxy apertures.}
    \label{tab:groupgas}
    \begin{tabular}{cccc}
    \hline
        && Entire System & Intragroup Gas \\
        \hline
        WALLABY & $S$ (Jy km s$^{-1}$) & 27.98$\pm0.72$ & 11.35$\pm0.88$ \\
        &  $\log(M_{\rm HI}/$M$_{\odot})$ & $10.17\pm0.01$ & $9.78\pm0.03$ \\
        \hline
        ATCA & $S$ (Jy km s$^{-1}$) & 27.07$\pm0.51$ & 12.87$\pm0.81$  \\
        & $\log(M_{\rm HI}/$M$_{\odot})$ & $10.16\pm0.01$ & $9.84\pm0.03$ \\
        \hline
    \end{tabular}
    
\end{table}

\begin{figure}
    \centering
    \includegraphics[width=0.49\textwidth]{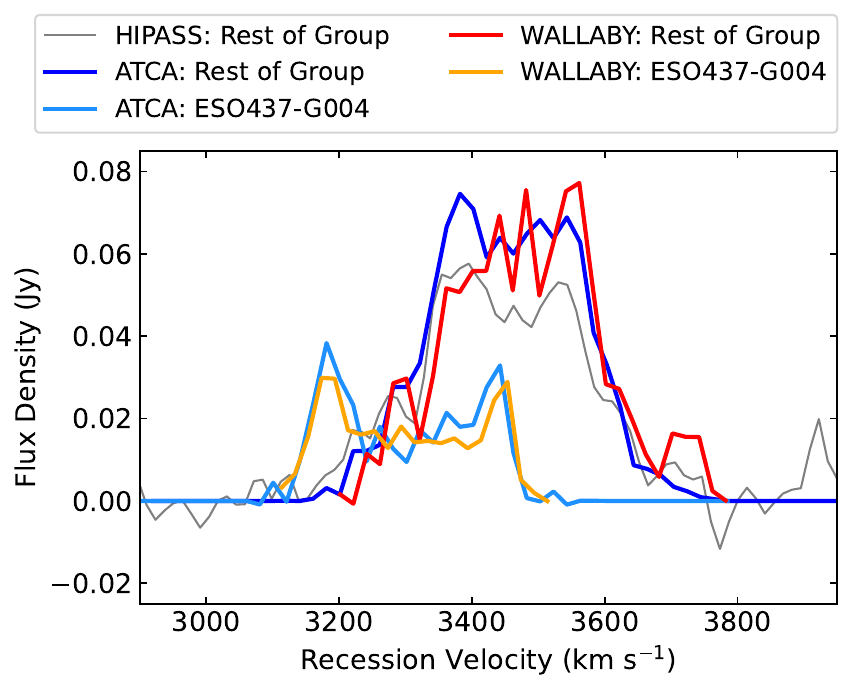}
    \caption{A comparison of the H{\sc i} spectra for ESO~437-G004 and the rest of the Klemola~13 system for the ATCA, WALLABY and HIPASS data. }
    \label{fig:spec}
\end{figure}

The H{\sc i} spectra for the Klemola~13 system are shown in Figure~\ref{fig:spec} for both the SoFiA masked WALLABY and ATCA data. Because the northern galaxy ESO~437-G004 has a distinct spatial separation from the rest of the system its spectra have been shown separately in orange and light blue. The WALLABY spectra have been smoothed to the same velocity resolution as the ATCA data (20 km s$^{-1}$) for comparison. It can be seen that the WALLABY and ATCA spectra span roughly the same width and there are some discrepancies. The spectrum that we measure from HIPASS data is also shown. This spectrum was measured using the {\tt MBSPECT} task in {\tt MIRIAD}, integrating over a $20\arcmin$ box centred on RA $=$ 10:35:49 and DEC $= -$28:31:51, sufficient to contain the intragroup gas seen in ATCA and WALLABY data and all of the galaxies except ESO~437-G004. The HIPASS data has a velocity resolution of 13.4 km s$^{-1}$ and a 15.5\arcmin ~beam. However, HIPASS cubes are optimised for point sources, and spatial filtering of extended sources may result in the loss of flux \citep{Barnes:01,Meyer:04}. We measure the integrated flux of the HIPASS spectrum to be 16.0 Jy km s$^{-1}$.

\begin{figure*}
     \centering
  
     \begin{subfigure}[b]{0.49\textwidth}
         \centering
         \includegraphics[width=\textwidth]{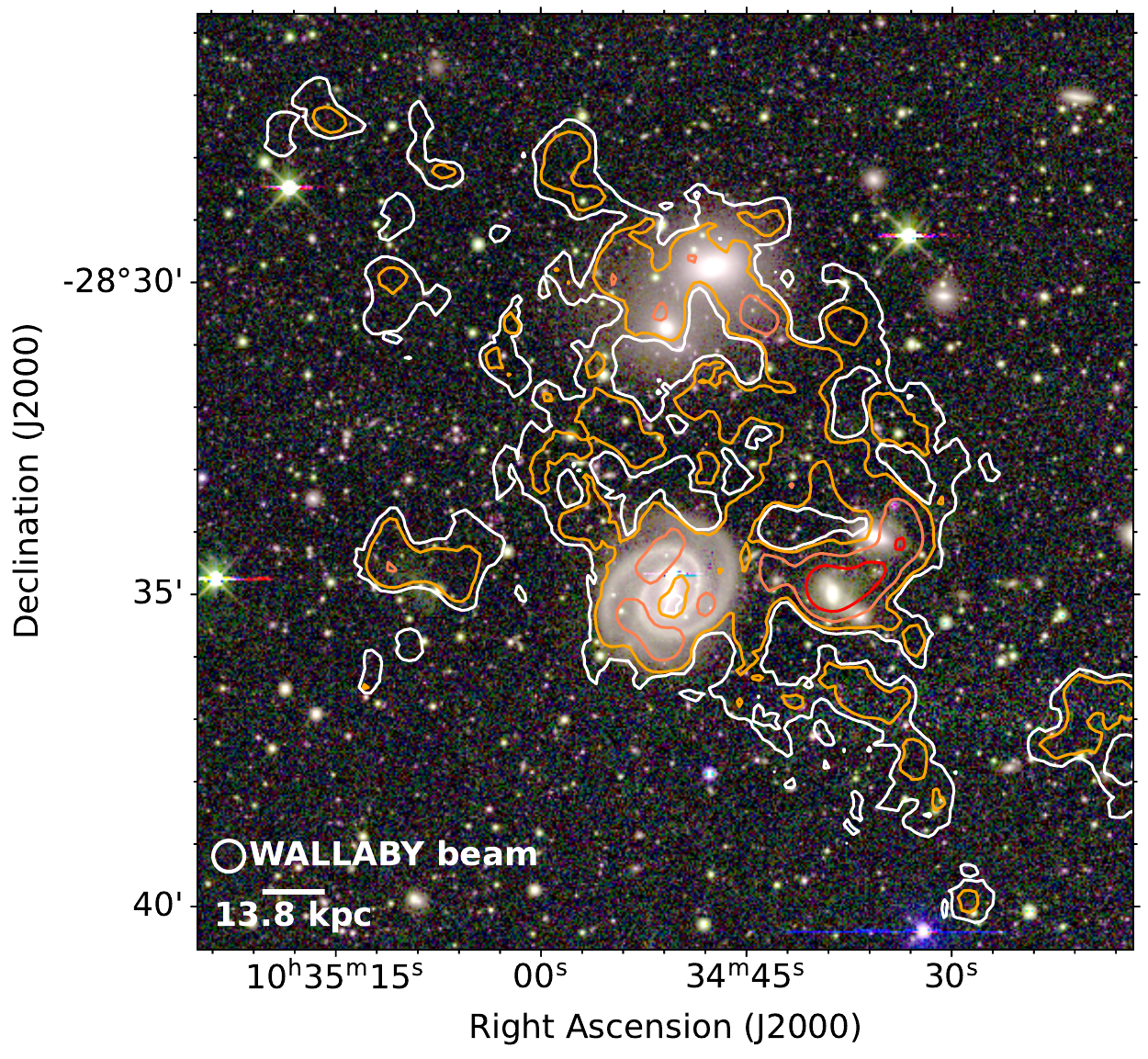}
         \caption{ }
         \label{fig:mom0W}
     \end{subfigure}
     \hfill
     \begin{subfigure}[b]{0.49\textwidth}
         \centering
         \includegraphics[width=\textwidth]{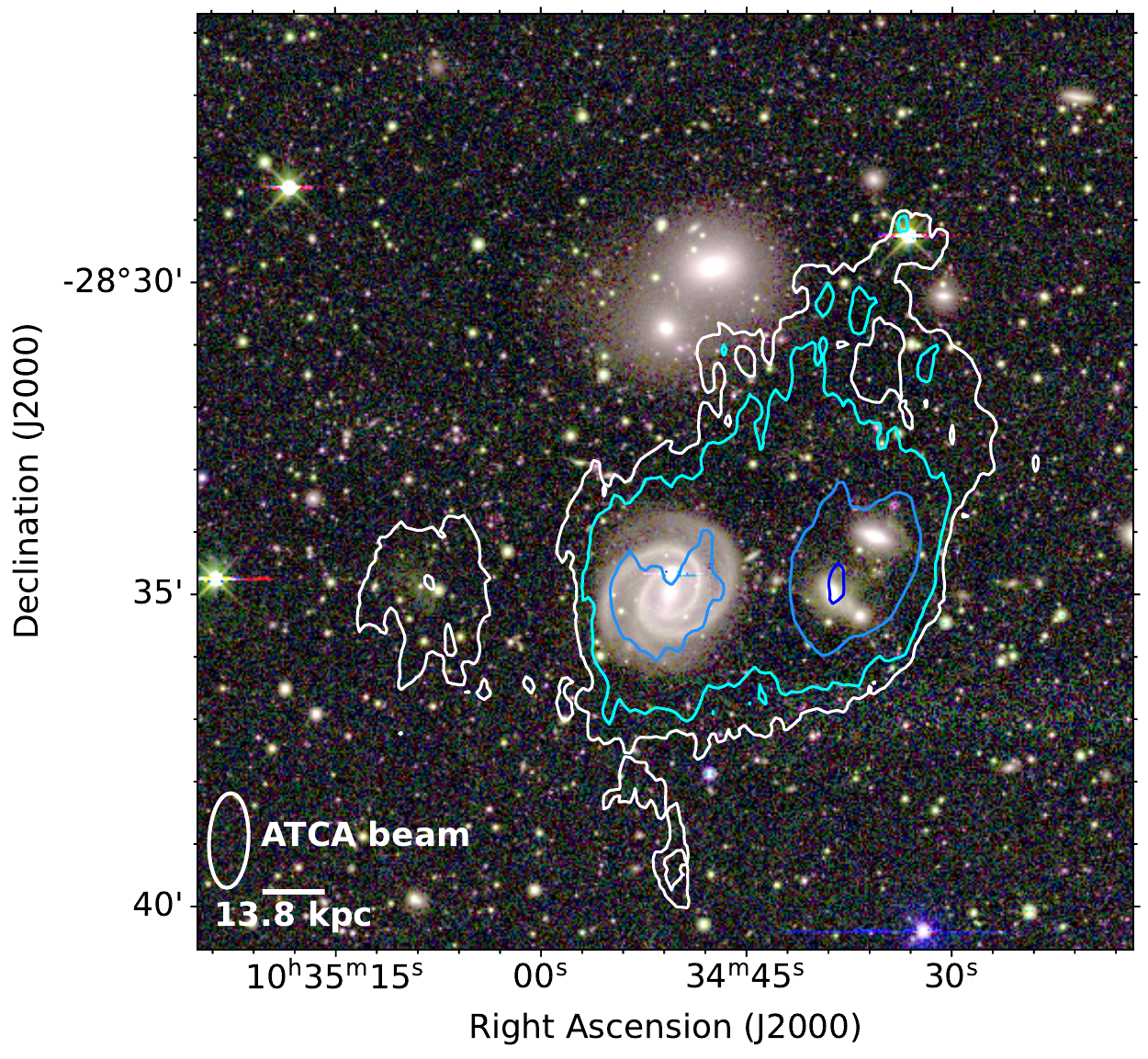}
         \caption{ }
         \label{fig:mom0A}
     \end{subfigure}

     \begin{subfigure}[b]{0.32\textwidth}
         \centering
         \includegraphics[width=\textwidth]{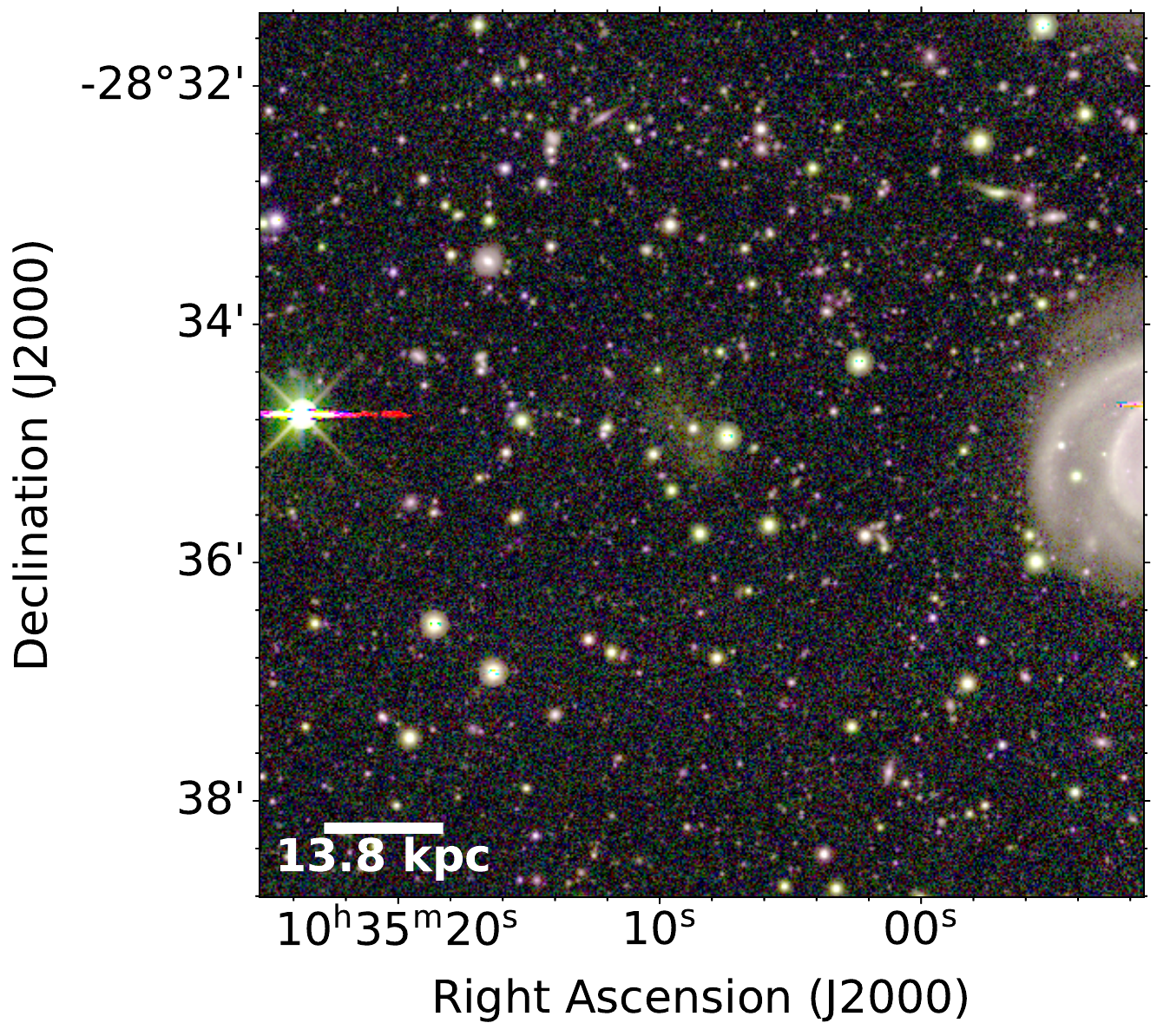}
         \caption{ }
         \label{fig:cloud}
     \end{subfigure}
     \begin{subfigure}[b]{0.32\textwidth}
         \centering
         \includegraphics[width=\textwidth]{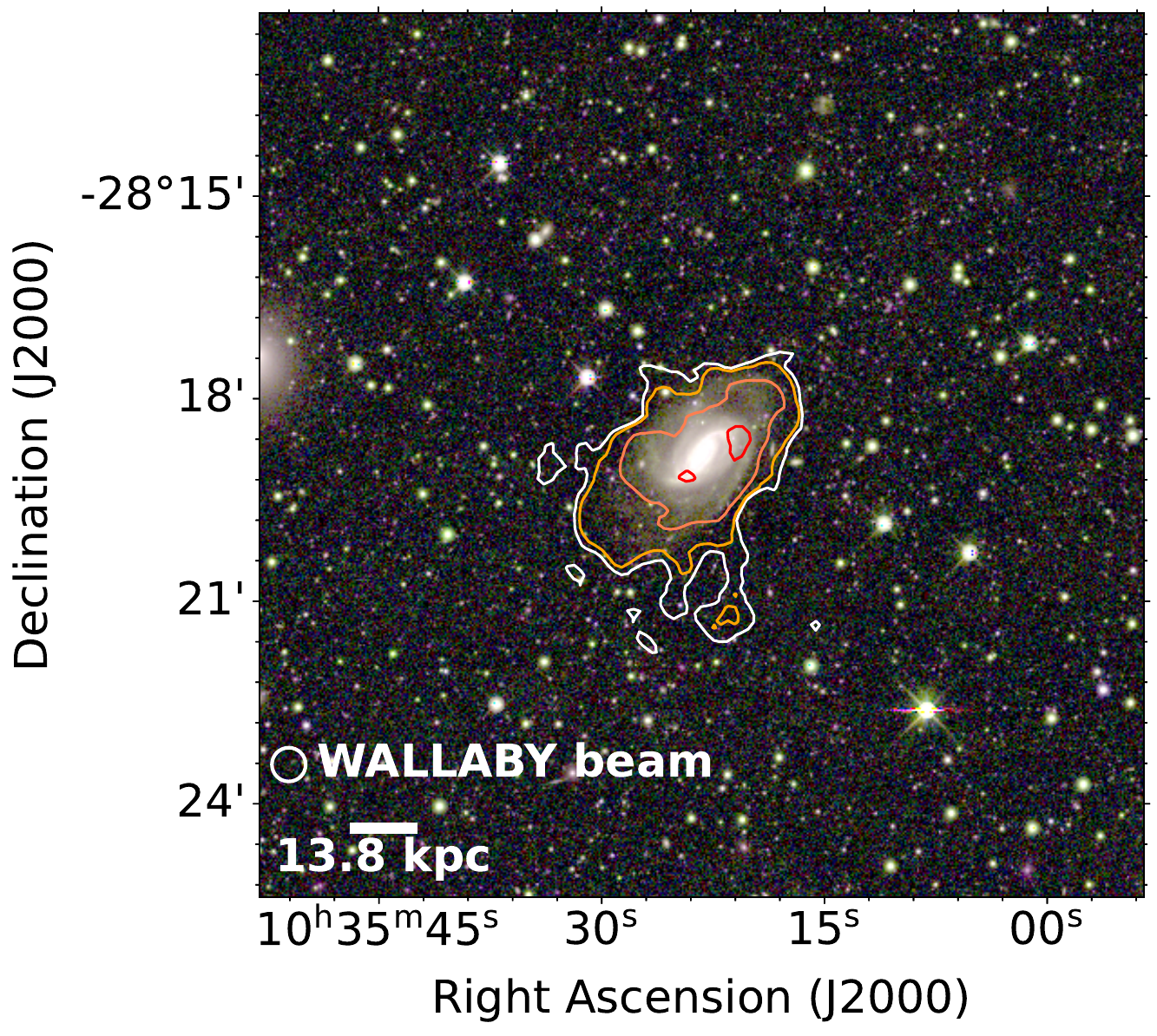}
         \caption{ }
         \label{fig:blobW_m0}
     \end{subfigure}
     \begin{subfigure}[b]{0.32\textwidth}
         \centering
         \includegraphics[width=\textwidth]{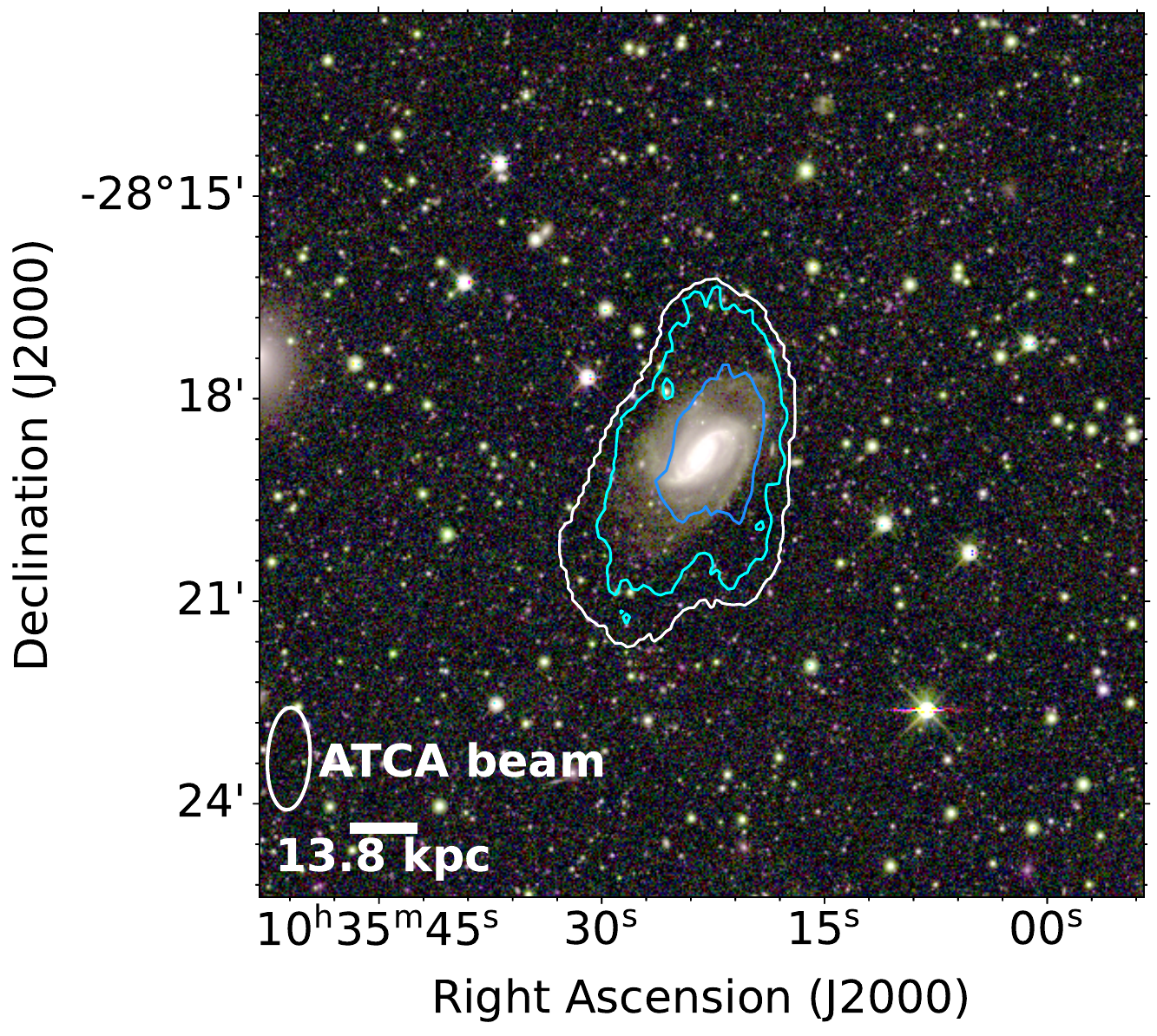}
         \caption{ }
         \label{fig:blobA_m0}
     \end{subfigure}
     
        \caption{Legacy Survey DR10 $grz$ images overlayed with H{\sc i} contours. The contour levels are $1\times10^{19}$ (white),  $1\times10^{20}$ (orange/cyan), $3\times10^{20}$ (coral/light blue), and $7\times10^{20}$ (red/dark blue) cm$^{-2}$.  (a) WALLABY contours over central region of the group. (b) ATCA contours over the central region of the group. (c) The Legacy Survey image centred on the almost dark cloud. (d) The WALLABY contours over ESO 347-G004. (e) ATCA contours over ESO 437-G004.}
        \label{fig:DESI_comp}
\end{figure*}

Figure \ref{fig:DESI_comp} shows the Legacy Survey DR10 $grz$ images with H{\sc i} column density contours from WALLABY and ATCA overlayed. The lowest contour levels in Figures \ref{fig:labels} and \ref{fig:DESI_comp} represent the emission at the edge of the SoFiA masks. The densest H{\sc i} regions line up well in the WALLABY and ATCA observations. In both Figures \ref{fig:mom0W} and \ref{fig:mom0A}, it can be seen that the highest density H{\sc i} is around PGC~031288, ESO-LV~4360421 and ESO~436-IG042, with a peak flux of $0.042/0.015$ Jy km s$^{-1}$ (WALLABY/ATCA) in the latter. The low column density contours can be seen stretching towards the northern galaxy, ESO~437-G004. WALLABY detects a higher column density of H{\sc i} in the almost dark cloud H1032-2819, while ATCA detects more of its extended emission. This is likely due to excellent coverage of short baselines in the ATCA observations and WALLABY's flux deficit issue that faint sources are prone to (see Section 6 in \citet{Westmeier:22}). 


\subsection{Comparison with the Literature}

\begin{figure}
    \centering
    \includegraphics[width=0.49\textwidth]{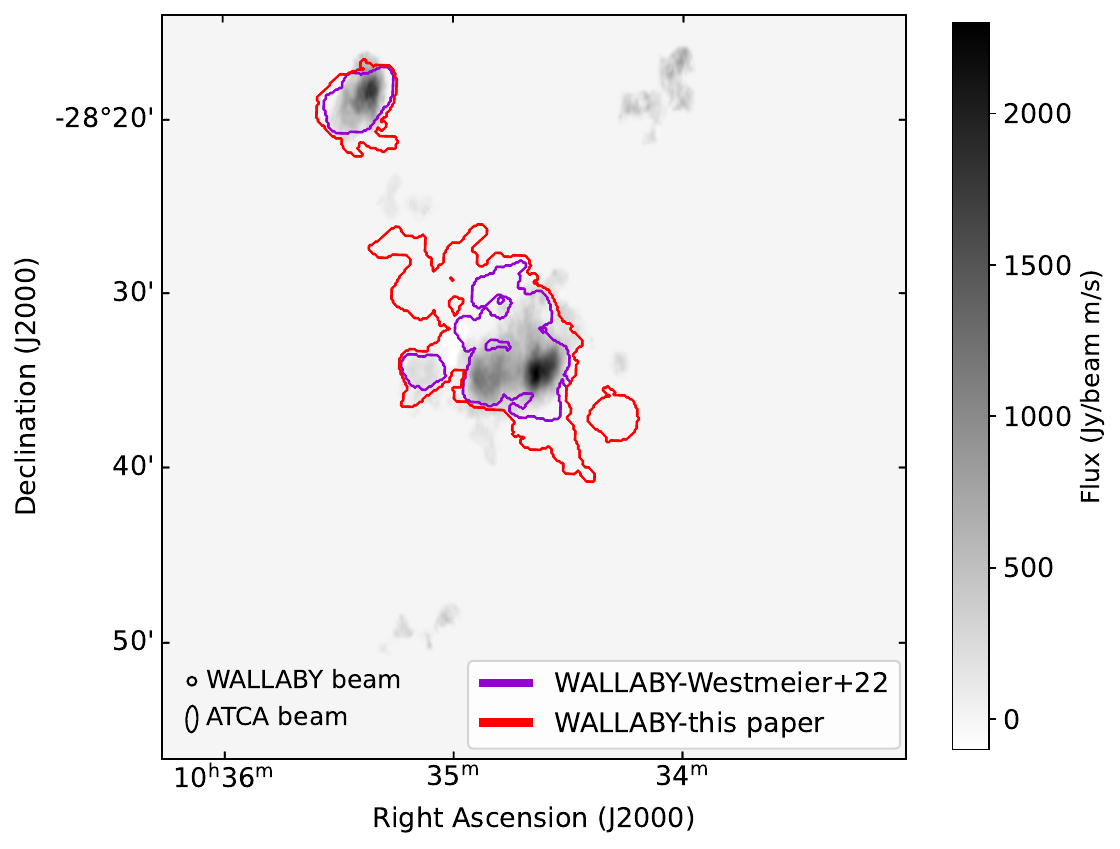}
    \caption{The 2D mask outlines of both WALLABY data sets overlayed on the ATCA column density image. In purple are the mask outlines for the sources as published in Westmeier et al. 2022 (WALLABY J103508-283427, WALLABY J103523-281855 and WALLABY J103442-283406), and in red are the mask outlines used in this study.} 
    \label{fig:masks}
\end{figure}

The masks differ significantly between the public release WALLABY data \citep{Westmeier:22} and the version presented here. Re-running SoFiA with parameters specific to this source rather than relying on the automated pipeline allows us to recover more extended emission. The automated pipeline is good for general WALLABY detections, however in diffuse, interacting regions such as this, varying the SoFiA parameters can allow for a more comprehensive analysis, as shown in Batten et al. (in preparation). Figure \ref{fig:masks} shows the outline of the flattened masks of both WALLABY data sets overlayed onto the ATCA column density image. Only the published WALLABY SoFiA run identifies H1032-2819 as a separate source distinct from the rest of Klemola~13. The WALLABY mask created for this paper suggests a bridge, which was not identified by \cite{McMahon:93} or \cite{Duc:99}.  Nevertheless, the velocity fields in Figure~\ref{fig:mom1} do show a discontinuity, implying that H1032-2819 is more probably a distinct source, that is seen in projection, rather than a simple easterly extension from ESO~436-G046.

To compare our measured properties of H1032-2819 with the previous detections, the published values of flux, central velocity and velocity width are shown in Table \ref{tab:res_pub}. The VLA data is from \cite{McMahon:93}, and the Nançay data is from \cite{Duc:99}. The WALLABY measurement using the mask defined in this paper has recovered the most flux. All of the central velocities lie within $\sim 50$ km s$^{-1}$ of each other. 

\begin{table}
    \centering
    \caption{ Comparison of the almost dark cloud with the literature. Uncertainties are included if given in the literature. Width refers to $w_{50}$ except $^{*}$ which indicates the width used in McMahon (1993) which is roughly equivalent to $w_{20}$.}
    \label{tab:res_pub}
    \setlength{\tabcolsep}{3.5pt}
    \begin{tabular}{cccc}
    \hline
         & S (Jy km s$^{-1}$) & $v_{\rm c}$ (km s$^{-1}$) & width (km s$^{-1}$)  \\
         \hline
         ATCA & $0.92\pm0.11$ & $3266\pm10$ & $57\pm20$ \\
         WALLABY-this paper & $1.37\pm0.14$ & $3255\pm2$ & $46\pm4$ \\
         WALLABY-Westmeier+22 & $0.70\pm0.09$ & 3295 & 135 \\
         Nançay & 0.96 & 3280 & 95 \\
         VLA & $0.44\pm0.10$ & 3304 & $\le 42^*$ \\
    \hline
    \end{tabular}
    
\end{table}

In both of the previous studies containing H1032-2819 \citep{McMahon:93,Duc:99}, this source is identified as a H{\sc i} cloud without an optical counterpart. However, using the most recent Legacy Survey DR10, for the first time an extremely faint optical counterpart is visible by eye. Section \ref{sec:mstar} explores the optical properties we have derived for this almost dark cloud.

\subsection{Stellar Mass and Specific Star Formation Rate}
\label{sec:mstar}

The stellar masses for all the galaxies except H1032-2819 were estimated from the Wide-field Infrared Survey Explorer \citep[\textit{WISE}][]{WISE:10} . This is done using the \textit{WISE} W1 and W2 magnitudes, as shown in Equations \ref{eq:Mstar} and \ref{eq:Mstar2} \citep{jarrett:23}: 
\begin{equation}
    \log \left( \frac{\Upsilon^{3.4\mu \textrm{m}}_{*}}{\textrm{M}_{\odot}/\textrm{L}_{\odot}} \right) = -0.376 - 1.053\times (W1-W2) ,
    \label{eq:Mstar}
\end{equation}
\begin{equation}
    \frac{L_{3.4\mu \textrm{m}}}{\textrm{L}_{\odot}} = 10^{-0.4(M-M_{SUN})} ,
    \label{eq:Mstar2}
\end{equation}
 where $\Upsilon^{3.4\mu \textrm{m}}_{*}$ is the stellar mass to light ratio in the W1 band, $L_{3.4\mu \textrm{m}}$ is the luminosity in the W1 band, $M$ is the absolute magnitude of the source in the W1 band, and $M_{SUN}$ is the absolute magnitude of the sun in the W1 band \citep[$M_{SUN} = 3.24$ mag;][]{Jarrett:13}. The scatter in Equation \ref{eq:Mstar} is 0.11 dex. The stellar mass of ESO~436-IG042 and ESO-LV~4360421 are summed in Table \ref{tab:results}. 

H1032-2819 was not detected by \textit{WISE} and thus the \textit{WISE} W1 and W2 magnitudes were not available, so the stellar mass is derived from the Legacy Survey DR10 images. The apparent magnitude and average surface brightness measured in an elliptical aperture containing all the visible emission for each of the bands available ($g,r,i,z$) for H1032-2819 are given in Table \ref{tab:opt}. In making this measurement we masked out the significant foreground stars and removed the sky background. Using the $g$ and $r$ bands, we estimated the stellar mass using the results of \cite{Du:20}, which specifically relates to the $g-r$ colour of low surface brightness galaxies. We estimate a stellar mass of $\log(M_*/M_{\odot}) = 7.98$ and a stellar mass to light ratio of $\Upsilon_{*}=1$ M$_{\odot}/$L$_{\odot}$. For comparison, we also calculated the stellar mass using the relation determined by \cite{Robotham:20}, which uses the spectral energy distribution of ProSpect. This approach gives rise to a higher value of $\log(M_*/M_{\odot}) = 8.35$. 

To perform the structural analysis of the object we additionally performed a fit to a Sérsic model using the {\tt IMFIT} software \citep{Erwin:15}. For this we performed a masking of all sources external to the object, using a combination of the masking provided using {\tt SExtractor} \citep{Bertin:96} and manual masks. First, we use a sum image of the $g$ and $r$ bands (the deepest ones) to obtain values for position, position angle ($33\fdg8 \pm 1\fdg5$) and ellipticity ($0.42\pm0.02$), which are subsequently fixed in the individual $g$, $r$, $i$ and $z$ band fitting, providing the rest of the following parameters: Sérsic index $n = 0.89 \pm 0.06$, effective radius $r_{eff} = 26.9\arcsec \pm 0.8\arcsec$ (6 kpc), and central $g$-band surface brightness $\mu_{cent,g} = 26.3 \pm 0.3$ mag arcsec$^{-2}$.

\begin{table}
    \centering
    \caption{Optical properties of H1032-2819. The properties from top to bottom are apparent AB magnitude and mean surface brightness.}
    \label{tab:opt}
    \setlength{\tabcolsep}{2pt}
    \begin{tabular}{ccccc}
    \hline
       Property  & $g$-band & $r$-band & $i$-band & $z$-band \\
        \hline
        $m_{AB}$ (mag)  & $18.6\pm0.3$ & $18.1\pm0.2$ & $17.5\pm0.2$ & $17.9\pm0.3$ \\
        $\mu$ (mag arcsec$^{-2}$) & $27.0\pm0.3$ & $26.4\pm0.2$ & $25.9\pm0.2$ & $26.3\pm0.3$ \\
    \hline
    
    \end{tabular}
    
\end{table}


The specific star formation rate (sSFR) was calculated using the method outlined in \cite{Janowieki:17} and \cite{Reynolds:22}. This method is summarised in Equations \ref{eq:sSFR} to \ref{eq:SFR_MIR4}:
\begin{equation}
    {\rm sSFR} = ({\rm SFR}_{\rm NUV} + {\rm SFR}_{\rm MIR})/M_* ,
    \label{eq:sSFR}
\end{equation}
\begin{equation}
    {\rm SFR}_{\rm NUV}/({\rm M}_{\odot} {\rm yr}^{-1}) = 10^{-28.165} L_{\rm NUV} /({\rm erg \: Hz} ) ,
    \label{eq:SFR_NUV}
\end{equation}
\begin{equation}
    {\rm SFR}_{\rm W3}/({\rm M}_{\odot} {\rm yr}^{-1}) = 4.91 \times 10^{-10} (L_{\rm W3} - 0.201 L_{\rm W1} )/{\rm L}_{\odot} ,
    \label{eq:SFR_MIR3}
\end{equation}
\begin{equation}
    {\rm SFR}_{\rm W4}/({\rm M}_{\odot} {\rm yr}^{-1}) = 7.50 \times 10^{-10} (L_{\rm W4} - 0.044 L_{\rm W1} )/{\rm L}_{\odot} ,
    \label{eq:SFR_MIR4}
\end{equation}
where SFR$_{\rm NUV}$ is the near ultraviolet star formation rate derived from $L_{\rm NUV}$, the \textit{GALEX} NUV band luminosity \citep{GALEX:05}. SFR$_{\rm MIR}$ is the mid-infrared star formation rate derived $L_{\rm W4}$ (or $L_{\rm W3}$ if the galaxy was undetected in W4) with a correction from $L_{\rm W1}$ to remove the contributions of old stars, where $L_{\rm W1}$, $L_{\rm W3}$ and $L_{\rm W4}$ \textit{WISE} W1, W3 and W4 band luminosities. If this correction was larger than the W3 or W4 luminosity contributions, then the MIR contribution to the total SFR is assumed to be zero (this is the case for PGC~031288, ESO~436-G044 and ESO~436-G045). This method of calculating the mid-infrared SFR was chosen by \cite{Janowieki:17} to enable comparison with the extended \textit{GALEX} Arecibo SDSS Survey \citep[xGASS,][]{Catinella:18}, which we also show in Section \ref{sec:scalingrel}. \cite{Reynolds:22} found that for the Hydra Cluster, their SFR values had error $\le 0.1$ dex using this method, so an uncertainty of 0.1 dex was assumed for our SFR values.  The SFR for ESO~346-IG042 and ESO-LV~4360421 are summed in Table \ref{tab:results}.  H1032-2819 was not detected in \textit{WISE} or \textit{GALEX}, so a $3\sigma$ upper limit to the SFR was estimated from the \textit{GALEX} NUV image using the H{\sc i} aperture. All of the derived stellar masses and specific star formation rates are presented in Table \ref{tab:results}, and the quantities used to calculate them are presented in Table \ref{tab:mstarsfr}.

\subsection{Kinematic Analysis}
\label{sec:kin}

\begin{figure*}
     \centering
     \begin{subfigure}[t]{0.49\textwidth}
         \centering
         \includegraphics[width=\textwidth]{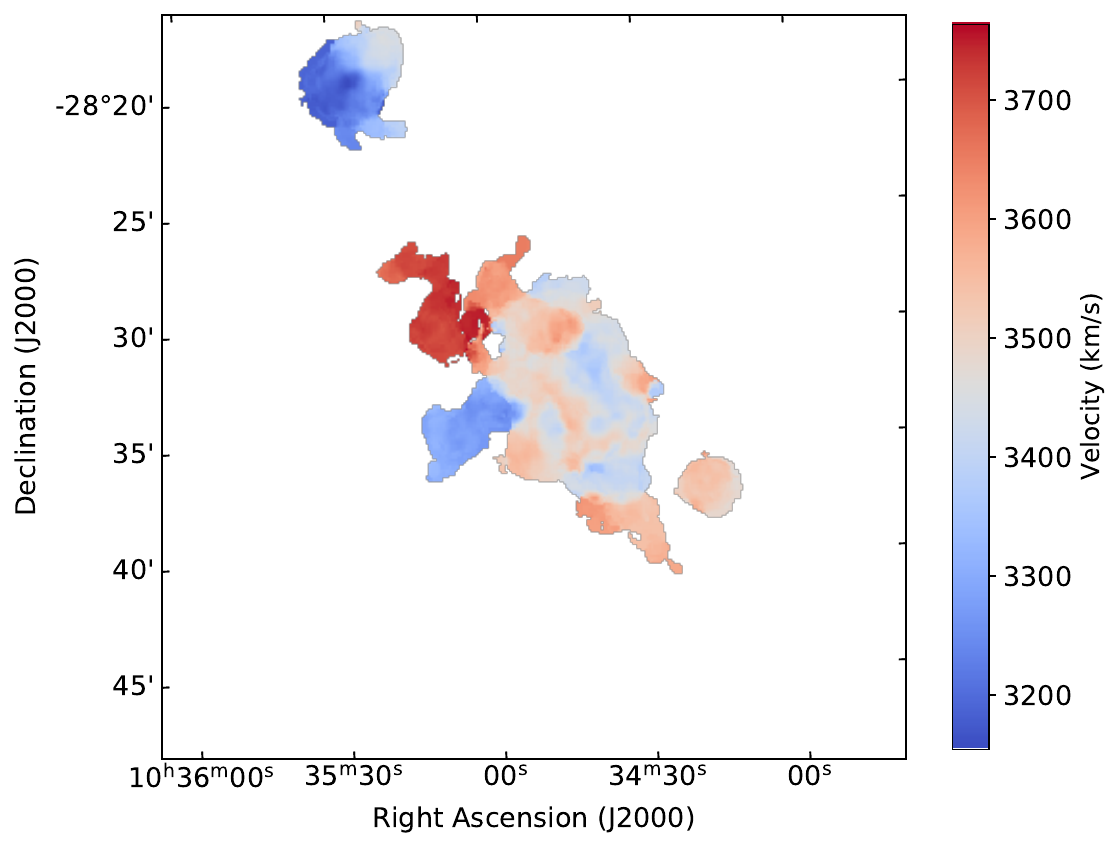}
         \caption{WALLABY System}
         \label{fig:mom1W}
     \end{subfigure}
     \hfill
     \begin{subfigure}[t]{0.49\textwidth}
         \centering
         \includegraphics[width=\textwidth]{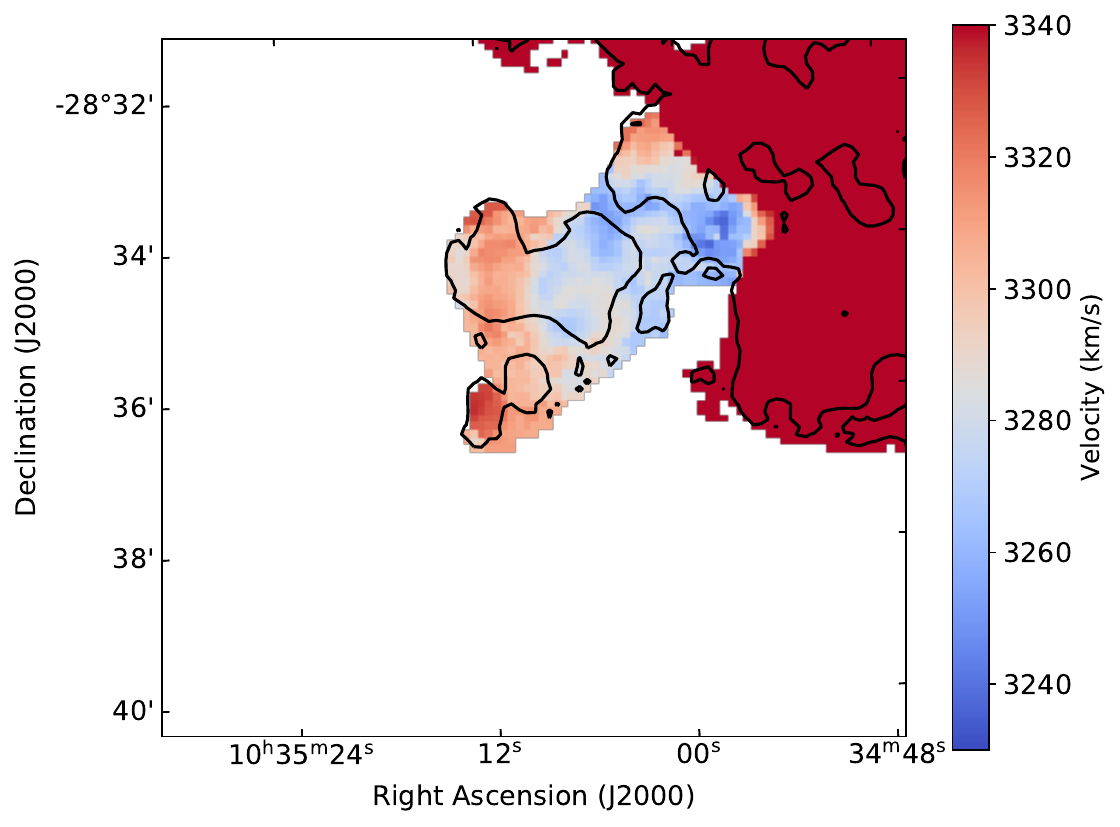}
         \caption{WALLABY H1032-2819}
         \label{fig:mom1W_D}
     \end{subfigure}
     \vspace{1cm}
     \begin{subfigure}[b]{0.49\textwidth}
         \centering
         \includegraphics[width=\textwidth]{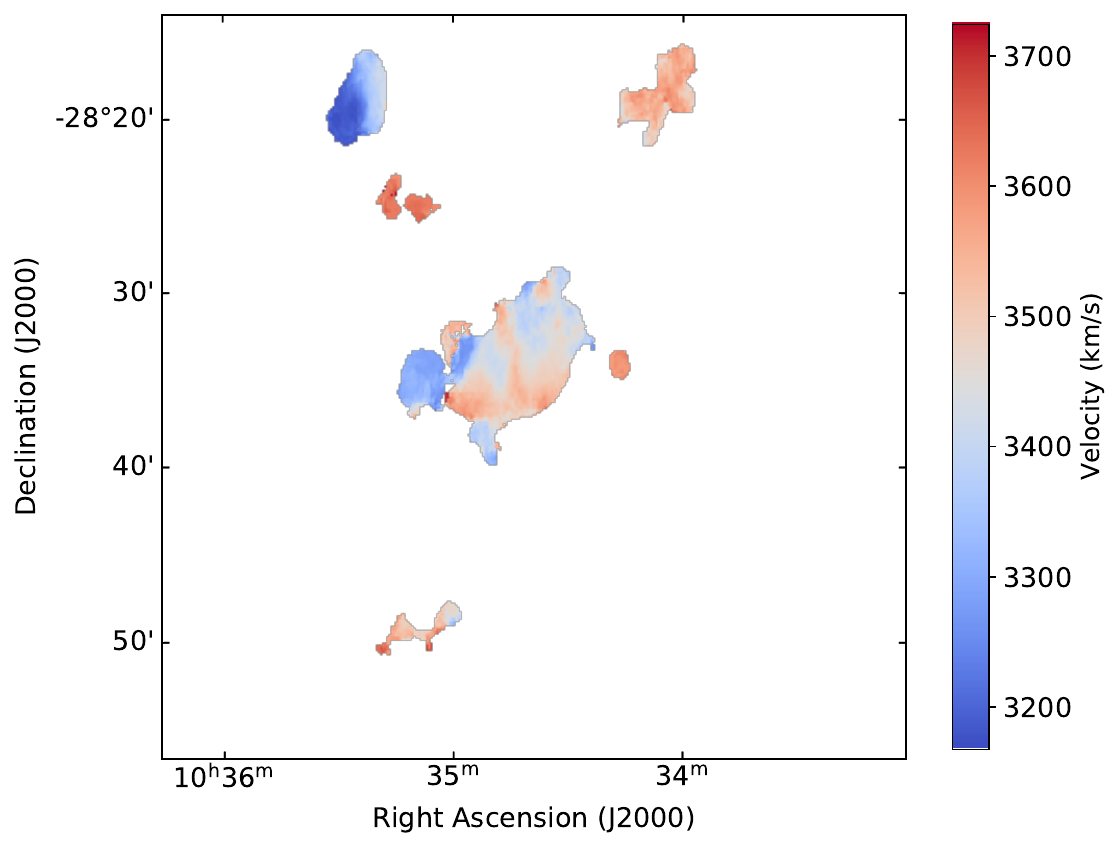}
         \caption{ATCA System}
         \label{fig:mom1A}
     \end{subfigure}
     \hfill
     \begin{subfigure}[b]{0.49\textwidth}
         \centering
         \includegraphics[width=\textwidth]{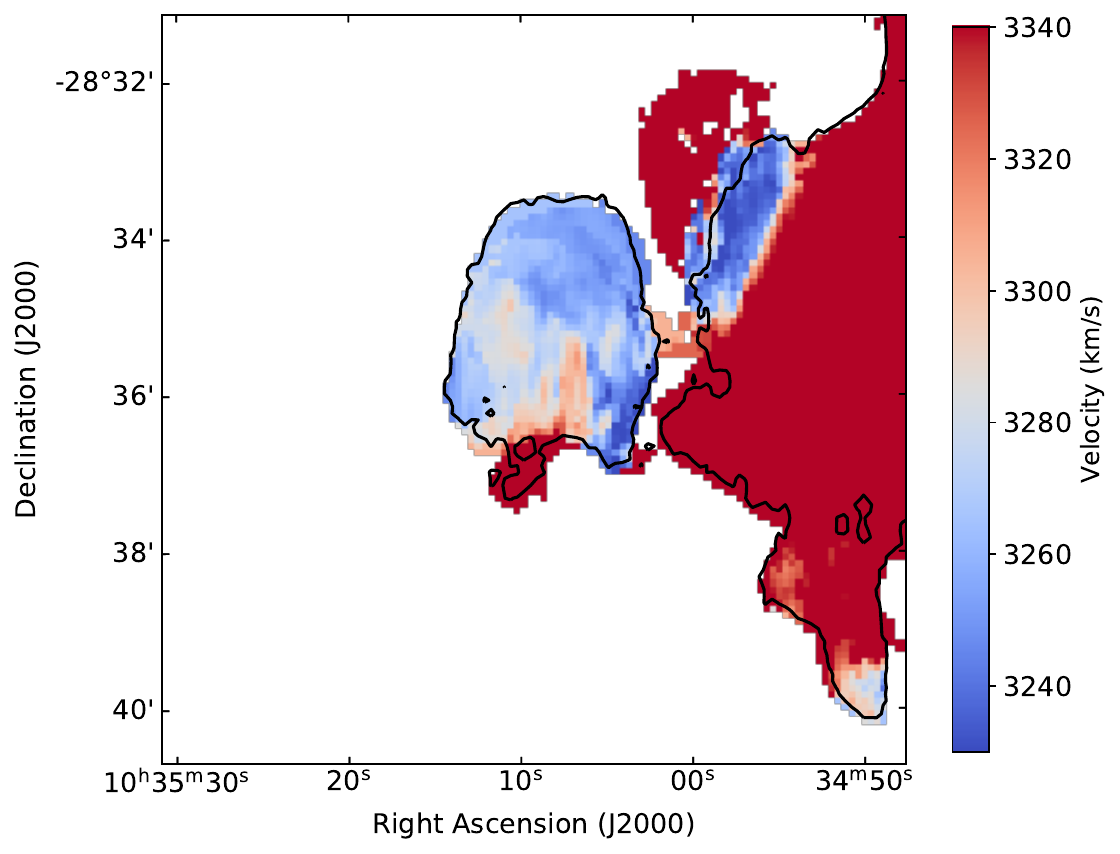}
         \caption{ATCA H1032-2819}
         \label{fig:mom1A_D}
     \end{subfigure}
        \caption{The H{\sc i} velocity fields from WALLABY (top row) and ATCA (bottom row). The left column is for the whole Klemola~13 system; the right column is zoomed in on the almost dark cloud H1032-2819. The black contour shows the $1\times10^{19}$cm$^{-2}$ column density level for the respective data sets.}
        \label{fig:mom1}
\end{figure*}

The velocity field for H1032-2819 can be seen in the moment 1 images in Figure~\ref{fig:mom1}. The mean velocity is similar to that of ESO~437-G004, but about 270 km~s$^{-1}$ lower than the more nearby ESO~436-G046. However, although the velocity width is $w_{50}=47$ km s$^{-1}$, it appears that H1032-2819 is not a regularly rotating disc. Common features between the WALLABY and ATCA velocity fields are that it is slightly more blueshifted to the west (towards ESO~436-G046), and slightly more redshifted in the south. 

The dynamical masses of all galaxies except ESO~436-G046 and ESO~437-G004 have been estimated using $M=rv^{2}/G$, where $v$ is half of the $w_{50}$ value, $r$ is the average of the semi major and minor axes of the aperture, and G is the gravitational constant. Sources with inclinations greater than $30^{\circ}$ have their $v$ value corrected for inclination by a factor of 1/sin$(i)$, where $i$ is the inclination calculated from the ratio of the major and minor axes of the optical apertures. Using this method, the dynamical mass of H1032-2819 is estimated to be $1\times10^{10}$ M$_{\odot}$ using an inclination of $50^{\circ}$. As this mass is $\sim 15 \times$ larger than the sum of the H{\sc i} and stellar mass, it is possible that H1032-2819 contains dark matter. However this is unlikely, as this estimate relies on the cloud being self gravitating in hydrostatic equilibrium. It is much more likely that tidal and ram pressure forces have contributed to a high mass estimate by stretching the source and adding velocity perturbations.

The more massive galaxies, ESO~437-G004 and ESO~436-G046, appear to be regular rotators, so {\tt 3DBarolo} \citep{DiTeodoro:15} was  used to model their kinematics. {\tt 3DBarolo} fits 3D tilted ring models to the H{\sc i} emission in the data cubes, resulting in the velocity fields and rotation curves shown in Figures \ref{fig:bbaroloblob} and \ref{fig:bbarolospiral}. For ESO~436-G046 the aperture shown in Figure~\ref{fig:labels} was applied to the SoFiA masked cube. For ESO~437-G004, we used the SoFiA mask that included its entire H{\sc i} content but only modelled out to a radius of $\sim20$ kpc, which excludes the outer regions as they are more affected by noise. {\tt 3DBarolo} was set to fit for three free parameters: rotation velocity ($V_{\rm rot}$), inclination ($i$) and position angle ($\phi$), as shown in Figures \ref{fig:pltblob} and \ref{fig:pltspiral}. We oversample the number of rings, fitting roughly two rings per beam, as done in the WALLABY kinematic modelling pipeline \citep{Deg:22} and in previous WALLABY studies \citep{Elagali:19,Reynolds:19}. The rotation velocities and dynamical masses derived from these models have been included in Table \ref{tab:results}, where the error in the dynamical mass is derived from the error in the rotation velocity. The optical coordinates from Table \ref{tab:results} were used as the centre, and a dispersion velocity of 11 km s$^{-1}$ was assumed, as typically found in massive and dwarf disc galaxies \citep[e.g.][]{Iorio:17,Bacchini:19,ManceraPina:21}. To estimate the rotation velocities for ESO~436-G046 and ESO~437-G004, the last 5 (4) and 4 (4) points in the WALLABY (ATCA) rotation curves were averaged respectively. Similar results for ESO~437-G004 were produced using both the ATCA and WALLABY data, which is a testament to the reliability of the fits. Meaningful models could not be created for all of the components of the system. The {\tt 3DBarolo} model for ESO~436-G046 is not reliable, as can be seen in the residuals in Figure~\ref{fig:m1spiral}, as it is not a perfectly regularly rotating disc. Nevertheless, we have included the rotation velocity derived from this model in Table \ref{tab:results} so that comparisons can be made. The kinematic models are quite sensitive to the initial input parameters, and smoothing over small scale structure can lead to improved residuals by a factor of $\sim3$ as described in Appendix \ref{sec:bbarolo}. 

Only ESO~437-G004 has a kinematic model in the WALLABY pilot data release 1 \citep[PDR1,][]{Deg:22}.  The WALLABY PDR1 kinematic model was created with the WALLABY Kinematic Analysis Proto-Pipeline (WKAPP). The WKAPP WALLABY model has an inclination of 58 degrees, a position angle of 316 degrees, and a rotation velocity of 163 km s$^{-1}$, which is higher than the values obtained here. WKAPP combines {\tt 3DBarolo} and {\tt FAT} \citep{Kamphuis:15} results, and does not allow the position angle and inclination to vary between the rings (as they do in our models). Although both produced rotation velocities higher than our models, the {\tt FAT} rotation velocity was significantly larger than that of {\tt 3DBarolo}. The other significant factor causing the increase in rotation velocity is the mask. In order to allow {\tt FAT} and {\tt 3DBarolo} run as similarly as possible, WKAPP creates its own masks. In contrast, we use the SoFiA mask to be consistent with the rest of our analysis. Allowing {\tt 3DBarolo} to create its own mask on our data gives higher outer rotation velocities using the ATCA data. For the WALLABY data, the entire rotation curve velocity is increased. However, this caused the overall kinematic model to have much higher residuals. 

\section{Discussion}
\label{sec:discuss}

\subsection{Global Galaxy Properties}
\label{sec:scalingrel}

\begin{figure*}
     \centering
     \begin{subfigure}[t]{0.46\textwidth}
         \centering
         \includegraphics[width=\textwidth]{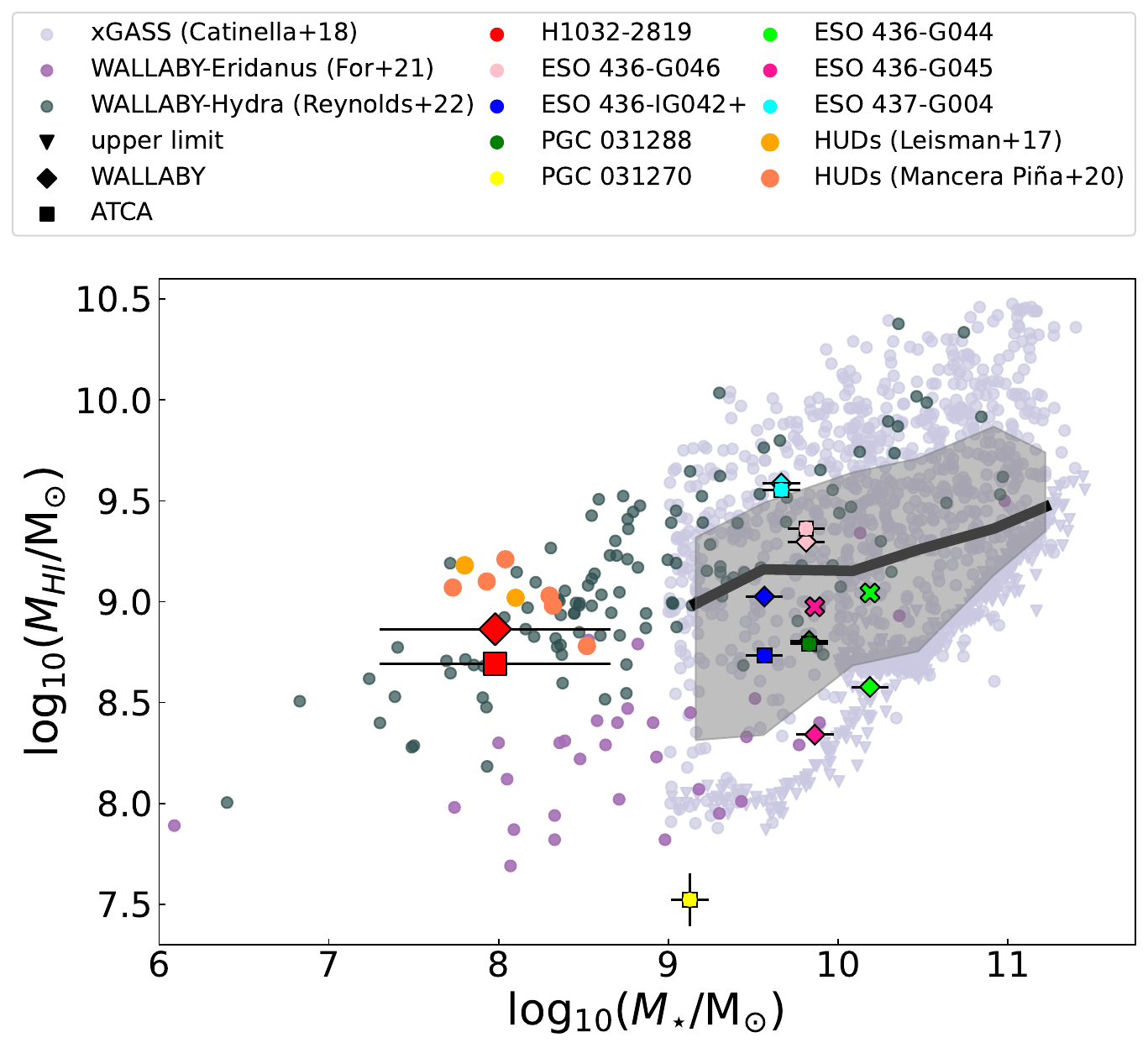}
         \caption{ }
         \label{fig:MHIstar}
     \end{subfigure}
     \begin{subfigure}[t]{0.5\textwidth}
         \centering
         \includegraphics[width=\textwidth]{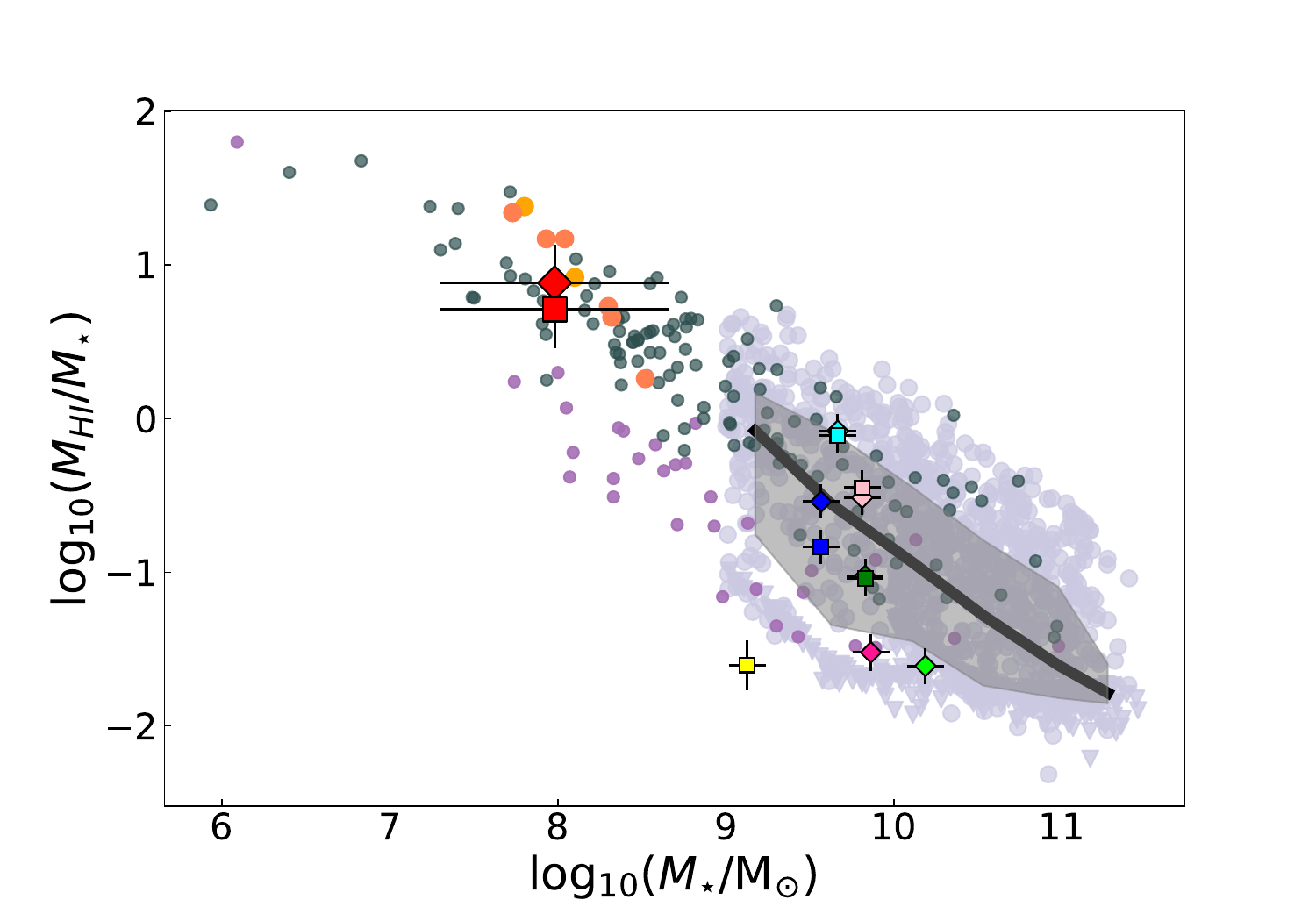}
         \caption{ }
         \label{fig:gfstar}
     \end{subfigure}
     \begin{subfigure}[b]{0.45\textwidth}
         \centering
         \includegraphics[width=\textwidth]{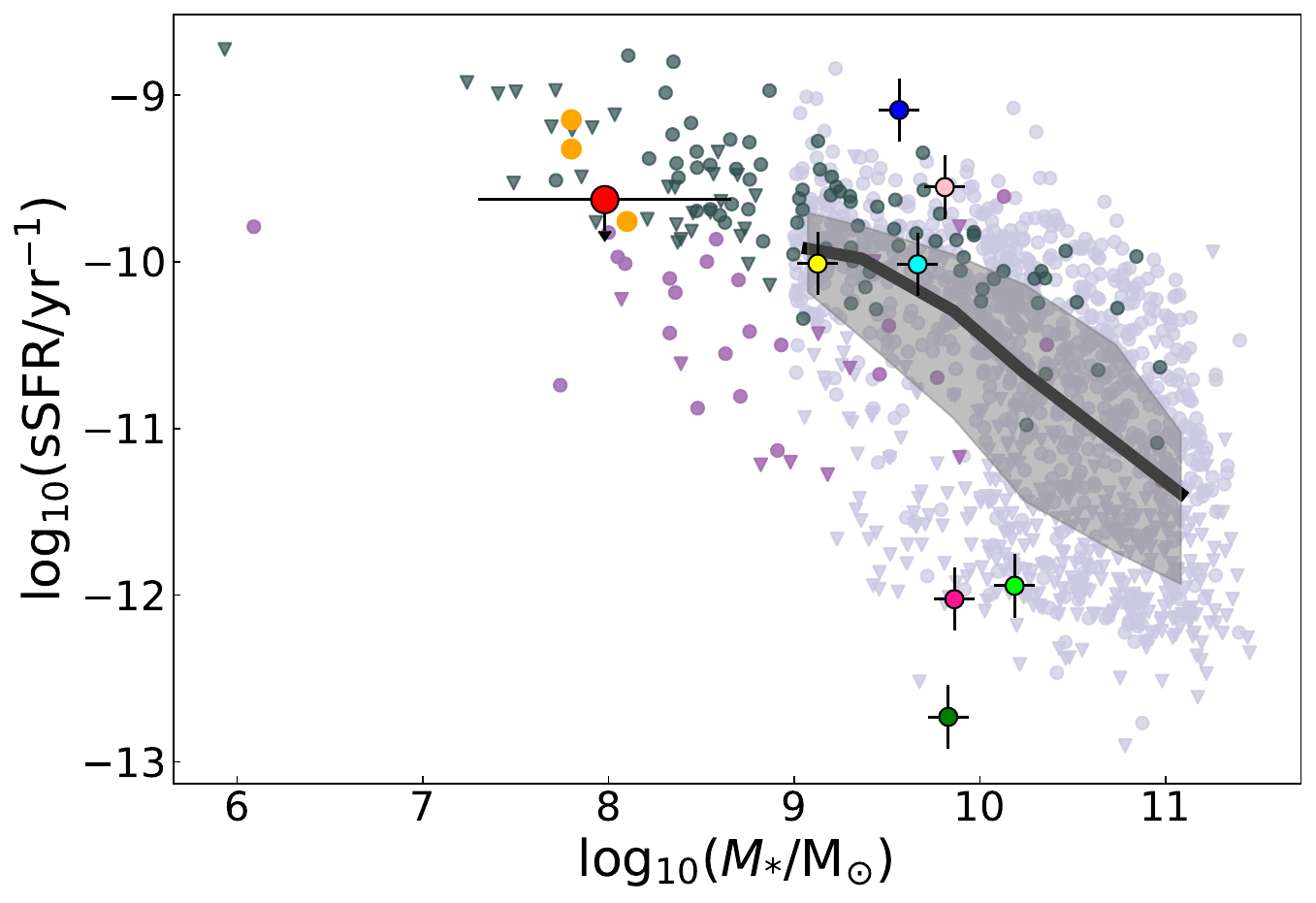}
         \caption{ }
         \label{fig:mstarSFR}
     \end{subfigure}
     \begin{subfigure}[b]{0.50\textwidth}
         \centering
         \includegraphics[width=\textwidth]{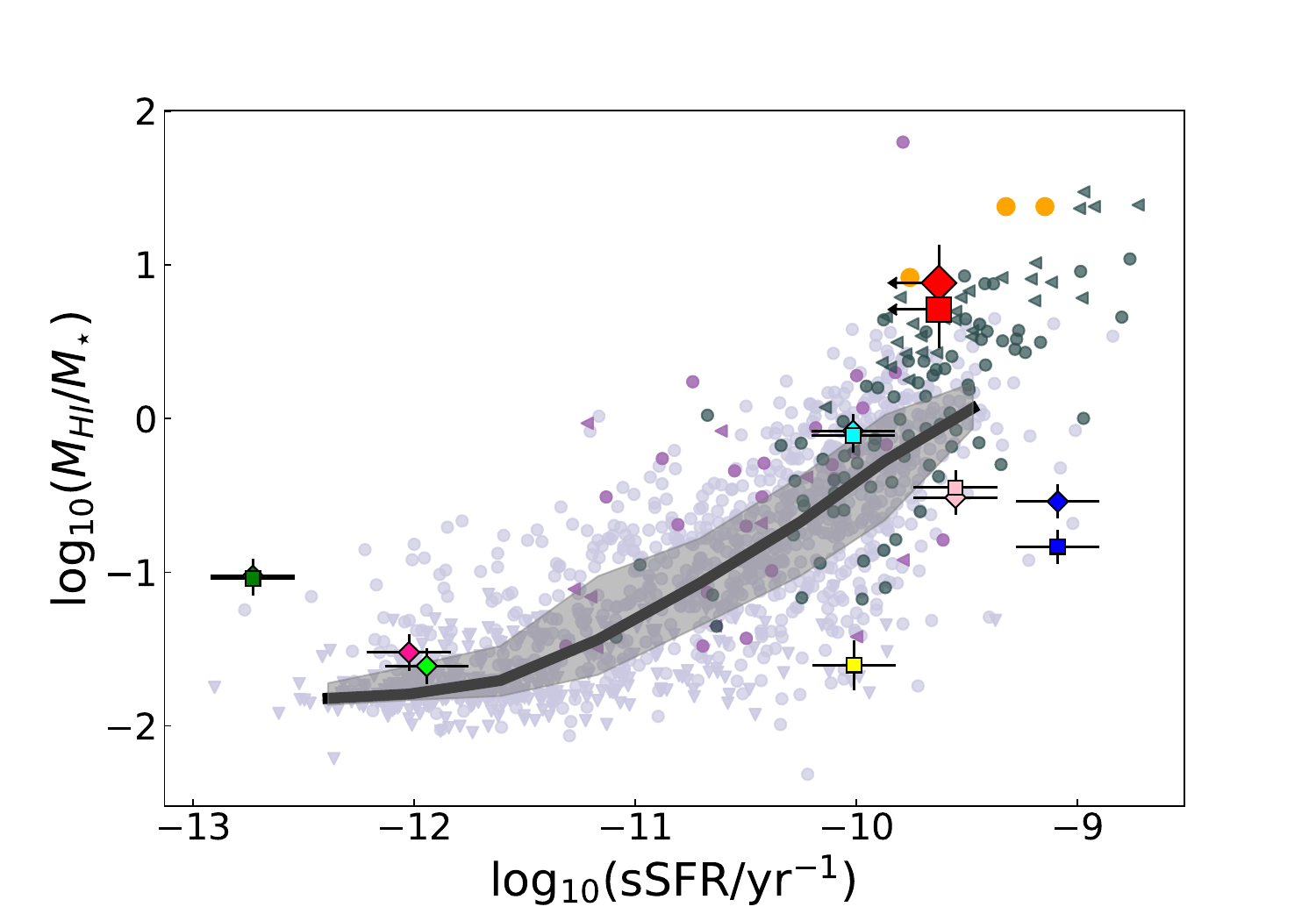}
         \caption{ }
         \label{fig:gfSFR}
     \end{subfigure}
        \caption{Scaling relations for the galaxies in Klemola~13 compared with xGASS (light grey), the WALLABY detections in the Hydra field (dark grey) and the WALLABY detections in the Eridanus supergroup (purple). HI-bearing UDGs have been included in orange. The solid line indicates the rolling median (with bin widths of 0.45) and the shaded region shows the interquartile region of the xGASS detections and non-detections. The almost dark cloud (H1032-2819, in red) has a larger marker for emphasis. The quantities derived for the sources in Klemola~13 can be found in Table \ref{tab:results}. In the legend, ESO~436-IG042+ refers to the aperture containing both ESO~436-IG042 and ESO-LV~4360421. (a) H{\sc i} mass against stellar mass. The cross symbols indicate that the H{\sc i} mass of the almost dark cloud has been added to the galaxy. (b) Gas fraction against stellar mass. (c) Stellar mass against specific star formation rate. (d) Gas fraction against specific star formation rate. }
        \label{fig:scale}
\end{figure*}

To see how the group environment affects the properties of H1032-2819 and the other galaxies in the interacting system, we have plotted them against galaxies from xGASS for common scaling relations in Figure~\ref{fig:scale}.  The WALLABY detections in the Eridanus supergroup \citep{For:21} and the WALLABY detections in the Hydra field \citep{Reynolds:22} are also plotted for comparison. HI-bearing UDGs from \cite{Leisman:17} and \cite{ManceraPina:20} have been included in orange. The stellar mass against H{\sc i} mass is plotted in Figure~\ref{fig:MHIstar} and the H{\sc i} gas fraction against stellar mass is plotted in Figure~\ref{fig:gfstar}. It is clear that H1032-2819 has a significantly lower stellar mass than all of the xGASS galaxies.  Only two galaxies lie above the median (ESO~436-G046 and ESO~437-G004). The three galaxies that appear extremely H{\sc i} deficient, ESO~436-G044, ESO~436-G045 and PGC~031270, have H{\sc i} masses below even that of scatter of the xGASS galaxies. Interestingly, adding the H{\sc i} mass of the almost dark cloud (H1032-2819) to the closer galaxies (ESO~436-G044 and ESO~436-G045) brings them up into the scatter of the xGASS galaxies, as shown by the cross markers in Figure~\ref{fig:MHIstar}. Hence, in general, the galaxies in this system tend to be H{\sc i} deficient. There are two explanations for this. Firstly, we are measuring the H{\sc i} gas within the apertures defined by each galaxy's optical radius, however it is common for galaxies to have H{\sc i} gas extending past twice the optical radius. If the apertures of the these galaxies are extended to twice the optical radii, PGC 031270 still has a H{\sc i} mass well below the scatter of the xGASS sample ($M_{HI} = 6.9\times10^7$ M$_{\odot}$), but the H{\sc i} masses of ESO 436-G044 and ESO 436-G045 are brought up just inside the scatter ($M_{HI} = 8.0\times10^8$ M$_{\odot}$ and $3.7\times10^8$ M$_{\odot}$ respectively). Secondly, the galaxies tend to be H{\sc i} deficient as the group environment has caused gas to be stripped from the individual galaxies and become shared between the components of the group. 

The stellar mass against specific star formation rate (sSFR) is shown in Figure~\ref{fig:mstarSFR} and the sSFR against gas fraction is shown in Figure~\ref{fig:gfSFR}. PGC~031288 has a lower sSFR than all the galaxies, apart from two xGASS sources. ESO~436-G046 is the only Klemola~13 galaxy with a sSFR above the median for its stellar mass. H1032-2819 has a higher gas fraction than all the xGASS galaxies.  The sSFR of H1032-2819 also sits at the high end of the range spanned by the xGASS galaxies, but for its given gas fraction, its sSFR is low. This is in line with the conclusions of \cite{KadoFong:22}, who find that UDGs tend to have low SFR as they are less efficient at converting H{\sc i} to molecular gas and to stars. 

\subsection{Origin and Evolution of the Almost Dark Cloud}

\subsubsection{Origin}

Due to the proximity of H1032-2819 to the other galaxies in Klemola~13 and its location near the Hydra Cluster, it appears unlikely that this almost dark cloud has a primordial origin. As discussed in Section \ref{sec:kin}, the cloud does not appear to be self-gravitating. However, there is insufficient data to definitively rule out a primordial origin, and there are no obvious tidal features in the optical. 

Instead, the cloud's origin and evolution is likely to be dominated by interactions. Uncovering the orbital history of each component of this system will not be possible given the observations available and the complexity of the system. Nevertheless we have devised two tidal fly-by interaction scenarios to give some insight into the likely formation mechanism of the almost dark cloud. In this case, in a tidal fly-by interaction the motion of the northern galaxy past the rest of the group causes gas to be stripped and form the cloud.

The first scenario considers that the almost dark cloud is the result of a fly-by interaction between the northern galaxy, ESO~437-G004, and one or both of the nearby extremely H{\sc i} deficient galaxies ESO~436-G044 and ESO~436-G045. It was shown previously in Figure~\ref{fig:MHIstar} that adding the H{\sc i} mass of H1032-2819 to either of these galaxies brings them back up to within the scatter of the xGASS sample. It is also important to note that the H1032-2819 and ESO~437-G004 have similar line-of-sight velocities. In this fly-by scenario, the gas from these less massive galaxies was removed by the tidal force of ESO~437-G004 and sling shot out of the group, forming the almost dark cloud in the foreground with a smaller recession velocity. 

Alternatively, the almost dark cloud could have formed through a tidal fly-by interaction with ESO~436-G046, the large spiral galaxy closest to its current position. In this case, the motion of ESO~437-G004 past ESO~436-G046, causes the H{\sc i} cloud to form from tidal stripping. To help quantify this, we calculated the tidal strength parameter, $S_{\rm tid}$, as given in \cite{Wang:22} for these two galaxies in relation to the cloud at their current positions: 
\begin{equation}
    S_{\rm tid}= \frac{M_{\rm gal}}{M_{\rm cloud}} \left (\frac{R_{\rm cloud}}{d_{\rm proj}}\right)^2 \left( \frac{v_{\rm circ,cloud}} {\Delta v_{\rm rad}} \right) ,
    \label{eq:Stid}
\end{equation}
where $M_{\rm gal}$ is the dynamical mass estimated using the {\tt 3DBarolo} model and $M_{\rm cloud}$ is the sum of the H{\sc i} and stellar masses of the cloud, $R_{\rm cloud}$ is the average of the major and minor radii of our defined aperture, $d_{\rm proj}$ is the projected distance between the cloud and the galaxy, $v_{\rm circ,cloud}$ is the circular velocity of the cloud (WALLABY value for self gravitating cloud), and $\Delta v_{\rm rad}$ is the difference between the line of sight velocities for the cloud and the galaxy (WALLABY values used). The projected distances between the cloud and ESO~436-G046, and the cloud and ESO~437-G004 are 54 kpc and 202 kpc respectively. Despite ESO~437-G004 having a much closer velocity to the cloud, the smaller projected distance of ESO~436-G046 leads to a tidal strength more than twice as large as that of ESO~437-G004 ($S_{\rm tid} = 3.4$ and 1.4 for ESO~436-G046 and ESO~437-G004, respectively). Of course, in this fly-by scenario, the formation of the cloud took place when ESO~437-G004 was much closer. If at some point in its orbital history ESO~437-G004 was the same distance from the cloud as ESO~436-G046 is today, ESO~437-G004 would have had a tidal strength $\sim8$ times larger ($S_{\rm tid} = 28$). This suggestion is supported by the velocity of the cloud being more similar to that of ESO~437-G004, rather than that of the more nearby galaxies. Using the relation between the half light radius and the virial radius from \cite{Kravtsov:13}, we also found that the dark cloud lies within the virial radii of both ESO~436-G046 and ESO~437-G004, making it especially susceptible to tidal forces. 

Like the tidal dwarf galaxies (TDG) in \cite{Kaviraj:12}, H1032-2819 has a stellar mass less than $10\%$ of its parents and lies within 15 optical half-light radii of the parent galaxies. It is also known than TDG do not contain a significant amount of dark matter \citep[e.g.,][]{Bournaud:06,Lelli:15,Gray:23}. Although it is possible that H1032-2819 may contain dark matter based of the dynamical mass estimation performed in Section \ref{sec:kin} (which relies on the assumption that it is self-gravitating), a tidal origin is still favoured. When estimating the dynamical masses for their tidal dwarf galaxies, \cite{Gray:23} found that including a rough correction for turbulence and asymmetric drift can significantly change the dynamical mass estimates.

H1032-2819 appears faint and diffuse, leading us to question whether the almost dark cloud could in fact be a UDG. The criteria defined by \cite{VanDokkum:15} is that a UDG must have an effective radius greater than 1.5 kpc and a central surface brightness fainter than 24 mag arcsec$^{-2}$. As discussed in Section \ref{sec:mstar}, the cloud has an effective radius of 6 kpc and a central $g$-band surface brightness of $26.3 \pm 0.3$ mag arcsec$^{-2}$. Thus the cloud does meet the criteria of a UDG, and bears similarities to HI-bearing UDGs \citep[HUDs, e.g.,][]{Leisman:17,ManceraPina:20}, which have been included in Figure~\ref{fig:scale}. UDGs can have a range of formation mechanisms, including tidal interactions, such as the UDGs studied in \cite{Jones:21}.


\subsubsection{Evolution}

\begin{figure*}
     \centering
     \begin{subfigure}[t]{0.46\textwidth}
         \centering
         \includegraphics[width=\textwidth]{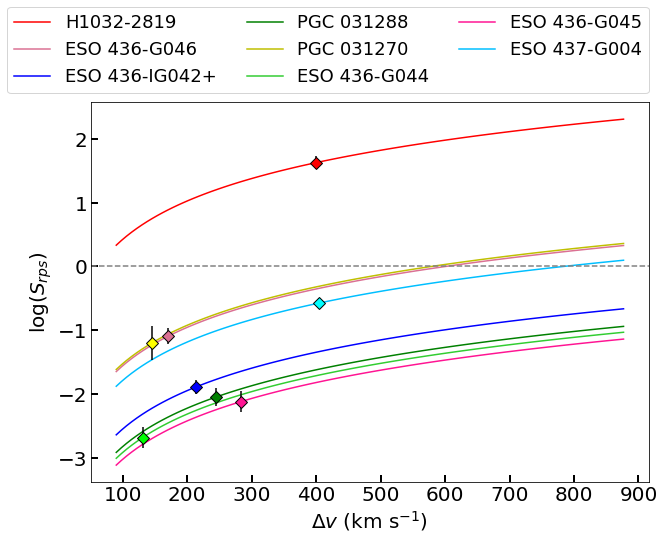}
         \caption{  }
         \label{fig:Srps_dv}
     \end{subfigure}
     \hfill
     \begin{subfigure}[t]{0.47\textwidth}
         \centering
         \includegraphics[width=\textwidth]{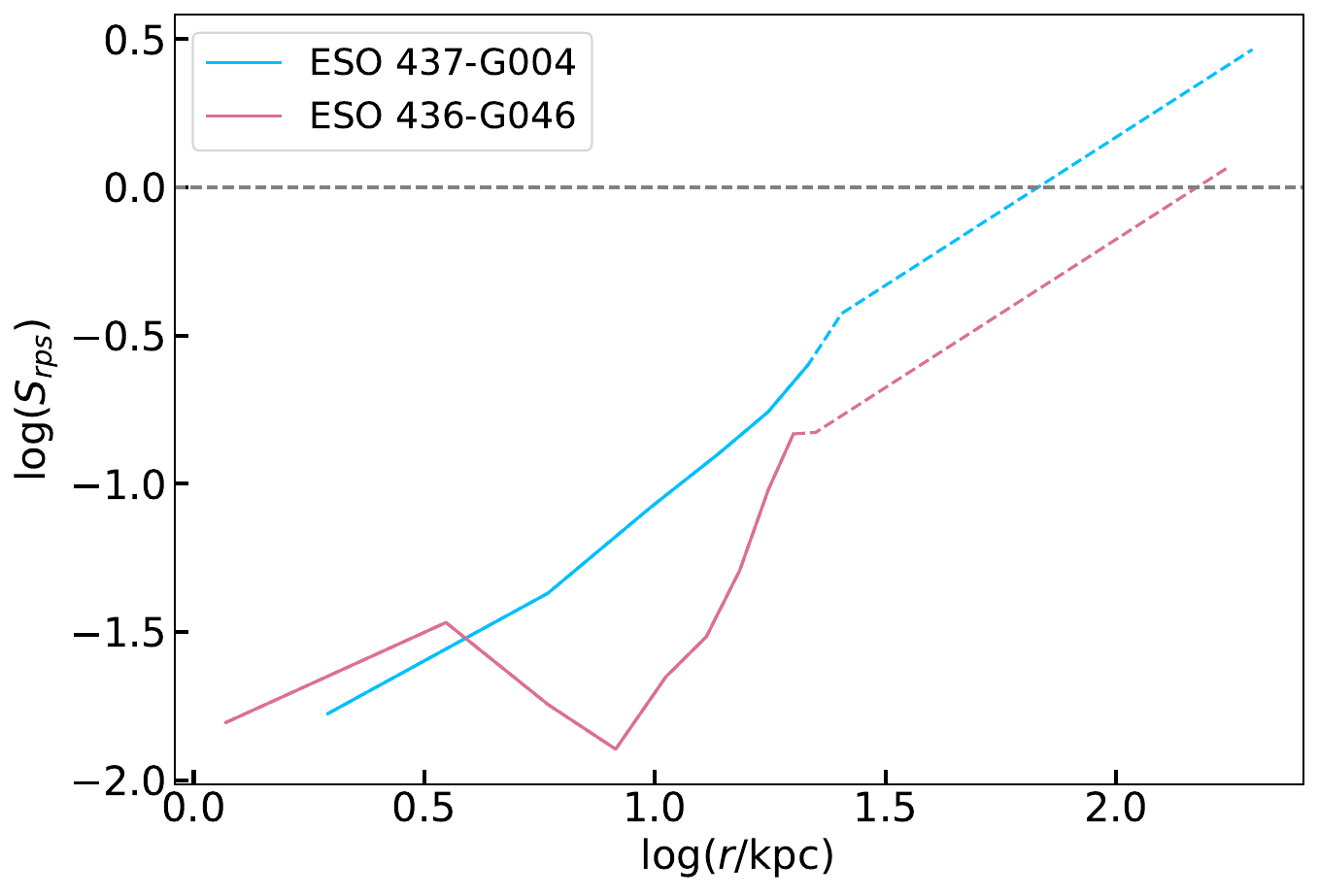}
         \caption{ }
         \label{fig:Srps-r}
     \end{subfigure}
        \caption{The ram pressure strength parameter (defined in Equations \ref{eq:Srps} to \ref{eq:Fanch2}) as a function of velocity difference from the cluster centre (left; a) and radius (right; b). Ram pressure becomes significant above the horizontal line at $S_{\rm rps}=1$.}
        \label{fig:Srps}
\end{figure*}

Being part of an interacting group, tidal forces can clearly play a dominant role in the origin of the almost dark cloud and in explaining the considerable amount of intragroup gas. But, due to the location of Klemola~13 just inside the virial radius of the Hydra cluster \citep[see][]{Reynolds:21}, ram pressure will also play an important part in the subsequent evolution of the gaseous components of the system and quenching of star formation in the group members.   
To quantify this effect, we have used the ram pressure stripping (RPS) strength parameter $S_{\rm rps}$, as defined in \cite{Lin:2023} and \cite{Wang:21}: 
\begin{equation}
    S_{\rm rps} = \frac{P_{\rm ram}}{F_{\rm anch}}
    \label{eq:Srps}
\end{equation}
\begin{equation}
    P_{\rm ram} = 1.4 m_{\rm p} n (\Delta v)^2
    \label{eq:Pram}
\end{equation}
\begin{equation}
    F_{\rm anch} = 2 \pi G (\Sigma_*+\Sigma_{\rm gas})\Sigma_{\rm gas} ,
    \label{eq:Fanch1}
\end{equation}
where values of $S_{\rm rps}$ greater than 1 indicate a susceptibility to ram pressure stripping. In this definition, $P_{\rm ram}$ is the ram pressure and $F_{\rm anch}$ is the anchoring force. $m_{\rm p}$ is the proton mass, $\Delta v$ is the galaxy velocity relative to the cluster medium, estimated using its line of sight velocity difference with the mean cluster velocity, $G$ is the gravitational constant, and $\Sigma_*$ and $\Sigma_{\rm gas}$ are the stellar and gas surface densities respectively. $n$ is intra-cluster medium number density at the projected distance from the cluster centre, as defined in \cite{Wang:21}. In our calculations we use the Hydra Cluster centre and velocity used in \cite{Wang:21} (RA = 159\fdg0865, Dec $= -27\fdg5629$ and $v = 3686$ km s$^{-1}$). For simplicity we assume a constant surface density when calculating $\Sigma_*$ and $\Sigma_{\rm gas}$ in Equation \ref{eq:Fanch1}. Errors associated with $M_*$ and $M_{\rm HI}$ will result in typical errors in $S_{\rm rps}$ of $\sim28\%$ and are shown in Figure ~\ref{fig:Srps_dv}. Figure~\ref{fig:Srps_dv} presents $S_{\rm rps}$ as a function of $\Delta v$ for each component of the system, with the markers representing the estimated values of $\Delta v$.\cite{Wang:21} argue that $n$ is an upper limit, and the estimated value of $\Delta v$ is a lower limit, and that the errors cancel out to some extent.  From Figure~\ref{fig:Srps_dv} we can see that the almost dark cloud is prone to ram pressure stripping regardless of the velocity difference from the cluster centre. Figure~\ref{fig:Srps-r} shows $S_{\rm rps}$ as a function of radius for ESO~436-G046 and ESO~437-G004. We calculate the value of $S_{\rm rps}$ in annuli with width $\Delta r$ equal to the value used in the respective {\tt 3D Barolo} models. For this plot, we include the combined effect of stars, gas and dark matter on the strength of the anchoring force. For Figure~\ref{fig:Srps-r} we replace Equation \ref{eq:Fanch1} with Equation \ref{eq:Fanch2} to calculate the anchoring force:
\begin{equation}
F_{\rm anch} = 2 \pi G \Sigma_{\rm dyn} \Sigma_{\rm gas} ,
\label{eq:Fanch2}
\end{equation}
where the rotation velocity and the gas surface density from {\tt 3DBarolo} at the given radius are used to calculate $\Sigma_{\rm dyn}$ and $\Sigma_{\rm gas}$ respectively. For details on the uncertainties associated with the {\tt 3D Barolo} models, see Section \ref{sec:kin} and Appendix \ref{sec:bbarolo}. The dotted lines in Figure~\ref{fig:Srps-r} extrapolate the trend beyond the radius of the {\tt 3DBarolo} models assuming a flat rotation curve (using the rotation velocity given in Table \ref{tab:results}) and a constant gas density equal to that of the outer radius of the model. This serves as an upper limit to $S_{\rm rps}$ as the outer radii beyond the model would have even lower gas surface densities. This figure highlights that the gas that remains today is tightly held, but extended gas that may have been associated to the galaxies in the past would have been susceptible to ram pressure stripping.


\subsection{Comparisons with Other H{\sc i} Clouds}

Klemola~13 has some similarities to other well-studied groups. Dark and almost dark clouds have been found to be rare by previous studies as they are difficult to identify in large numbers due to the relative surface brightness sensitivity limits in both the large H{\sc i} surveys and the large optical surveys. By combining Parkes and ATCA observations of the NGC 3783 group, \cite{kilborn:06} found one detection, GEMS\_N3783\_2, that did not have an optical counterpart in the NASA/IPAC Extragalactic Database (NED) or 6dFGS. It has an H{\sc i} mass of $4\times10^8$ M$_{\odot}$, comparable to the $7\times10^8$ M$_{\odot}$ almost dark cloud in Klemola~13. Similarly, they find that their dark cloud is most likely to be the result of a tidal interaction between the nearby gas rich spiral (ESO 378-G003) and one or more galaxies in the group. This is evident from the proximity of ESO 378-G003 to the dark cloud (a projected distance of 450 kpc) and its irregular H{\sc i} and velocity distribution on one side of the galaxy. Alternatively, they suggest that the dark cloud could be an extremely low surface brightness galaxy, with an upper limit of 22 mag arcsec$^{-2}$ in the $B$-band. In fact, some low surface brightness features can be seen in the Legacy Survey DR10 image.

The Leo I group is well known for the Leo Ring which contains $1.7\times10^9$ M$_{\odot}$ of H{\sc i} gas across a 225 kpc arc \citep{Schneider:89}. It only contains a few isolated patches of optical emission, so is considered dark on a large scale. \cite{Taylor:22} study five dark H{\sc i} clouds and one with a faint optical counterpart in this group with data from the Arecibo Galaxy Environment Survey \citep{Auld:06}. The sixth cloud has a smaller stellar mass ($M_{*} = 6.6\times10^6$ M$_{\odot}$) and H{\sc i} to stellar mass ratio ($M_{HI}/M_{*} = 1.4$) than our almost dark cloud. These clouds all have significantly smaller H{\sc i} masses and velocity widths than the one found in our system, with H{\sc i} masses in the range $2.5-9\times10^6$ M$_{\odot}$ and velocity widths in the range $15<w_{50}<42$ km s$^{-1}$. Although other formation mechanisms cannot be completely ruled out, like Klemola~13 and the NGC 3783 group, a tidal origin is favoured for all of the clouds. Three of the dark clouds are sandwiched between two spiral galaxies, M95 and M96, which have a projected separation of 136 kpc. Their proximity to the dark clouds, their similar velocities, and the warp seen in M96 suggests that the dark clouds are debris from a tidal interaction, despite there being no evidence of a tail.  


\cite{Cannon:15} presented VLA follow-up observations of five almost dark objects from ALFALFA. Using their own analysis of the SDSS images, rather than the results from the SDSS pipeline, they found that four of their sources did in fact have an extremely faint optical counterpart, like H1032-2819. However, H1032-2819 is so faint that the sensitivity of SDSS would be too low to detect it, and it can not be seen in the Digitized Sky Survey (DSS) image. Hence \cite{Cannon:15} acknowledge that it is possible that deeper imaging could reveal an optical counterpart to the fifth object. Due to its location in the NGC 3370 group, they conclude that this object is also the result of tidal interactions rather than primordial, while the other four galaxies are most likely to be `failed' galaxies with an extreme case of inefficient star formation. They have H{\sc i} to stellar mass ratios ranging from 14 to 75 and $g$-band H{\sc i} mass to light ratios ranging from 3 to 10, although \citet{Cannon:15} noted that  deeper imaging would likely decrease the mass to light ratios of the other four objects. With the combination of new large H{\sc i} and optical surveys such as WALLABY, the Legacy Survey archive and the Rubin Observatory \citep{Ivezi:19ApJ}, we have entered an era when we can begin to probe the low surface brightness universe in a much more systematic manner.

\section{Summary and Conclusions}
\label{sec:concl}

A combination of WALLABY and archival ATCA data \citep{LopezSanchez:08} has been used to perform an in-depth study of the almost dark cloud H1032-2819, and eight nearby galaxies (ESO~436-G046, ESO~436-IG042, ESO-LV~4360421, PGC~031288, PGC~031270, ESO~436-G044, ESO~436-G045 and ESO~437-G004) in the interacting group Klemola~13 (also known as HIPASS J1034-28 and the Tol 9 group). We analysed the H{\sc i} distribution as well as the kinematics of the system. H1032-2819 was first identified as a dark galaxy candidate in the VLA observations of \citet{McMahon:93} and confirmed in Nançay observations by \citet{Duc:99}. However, we have identified an extremely faint optical counterpart in data from the Legacy Survey DR10. The stellar mass-to-light ratio is $\Upsilon_{*} = 1.0$ M$_{\odot}$/L$_{\odot}$ and the estimated HI-to-stellar mass ratio is $M_{HI}/M_{*} = 7.6$ (using the WALLABY data). It has a mean $g$-band surface brightness of $27.0\pm0.3$ mag arcsec$^{-2}$.

Like other dark, or almost-dark, galaxy candidates, H1032-2819 is unlikely to have a primordial origin \citep[e.g.,][]{kilborn:06,Taylor:22}. The  proximity of H1032-2819 to other galaxies in Klemola~13 makes a tidal origin likely. We have outlined two fly-by scenarios that give plausible explanations for the formation of the  cloud. The fact that almost half of the H{\sc i} in Klemola~13 is contained in the intragroup medium, and that most of the member galaxies are H{\sc i} deficient compared to the xGASS sample, supports this scenario. 

The standard WALLABY source-finding pipeline (SoFiA) missed some extended gas in Klemola~13, but this was recovered with the use of bespoke SoFiA parameters which resulted in a larger mask. The peak column density is associated with ESO~436-IG042. WALLABY was able to detect H{\sc i} gas in two galaxies that ATCA did not (ESO~436-G044 and ESO~436-G045), although the ATCA data does show H{\sc i} gas reaching towards these galaxies. ATCA was able to detect H{\sc i} gas in one galaxy that WALLABY did not, PGC~031270.

The position of the cloud and Klemola~13 just inside the virial radius of Hydra Cluster suggests that ram pressure forces will probably drive future evolution, assuming that Klemola~13 is falling towards Hydra. As demonstrated by $S_{\rm rps}$ values consistently greater than 1, H1032-2819 and the intragroup gas are already susceptible to ram pressure stripping, although the gas within the massive galaxies ESO~436-G046 and ESO 347-G004 is sufficiently anchored. The latter two galaxies are in regular rotation, allowing kinematic models and total mass estimates were derived using {\tt 3DBarolo}.


This paper showcases the potential of the full WALLABY survey to detect H{\sc i} clouds and other extremely low surface brightness objects in the local Universe. WALLABY has a combination of column density sensitivity and angular resolution which allows easy cross-matching with optical/IR counterparts, even in the absence of spectroscopic data, and therefore easy identification of dark galaxy/cloud candidates. Further studies will concentrate on data from phase 2 of the pilot survey and the full WALLABY survey in combination with deep imaging data from the Legacy Survey. 


\section*{Acknowledgements}

We thank the anonymous referee for improving this manuscript.

This scientific work uses data obtained from Inyarrimanha Ilgari Bundara / the Murchison Radio-astronomy Observatory. We acknowledge the Wajarri Yamaji People as the Traditional Owners and native title holders of the Observatory site. CSIRO’s ASKAP radio telescope is part of the Australia Telescope National Facility (https://ror.org/05qajvd42). Operation of ASKAP is funded by the Australian Government with support from the National Collaborative Research Infrastructure Strategy. ASKAP uses the resources of the Pawsey Supercomputing Research Centre. Establishment of ASKAP, Inyarrimanha Ilgari Bundara, the CSIRO Murchison Radio-astronomy Observatory and the Pawsey Supercomputing Research Centre are initiatives of the Australian Government, with support from the Government of Western Australia and the Science and Industry Endowment Fund.

The Australia Telescope Compact Array is part of the Australia Telescope National Facility which is funded by the Australian Government for operation as a National Facility managed by CSIRO. We acknowledge the Gomeroi people as the traditional owners of the Observatory site. 

Parts of this research were supported by the Australian Research Council Centre of Excellence for All Sky Astrophysics in 3 Dimensions (ASTRO 3D), through project number CE170100013. 

This investigation has made use of the NASA/IPAC Extragalactic Database (NED) which is operated by the Jet Propulsion Laboratory, California Institute of Technology, under contract with the National Aeronautics and Space Administration, and NASA's Astrophysics Data System.

The Legacy Surveys consist of three individual and complementary projects: the Dark Energy Camera Legacy Survey (DECaLS; Proposal ID \#2014B-0404; PIs: David Schlegel and Arjun Dey), the Beijing-Arizona Sky Survey (BASS; NOAO Prop. ID \#2015A-0801; PIs: Zhou Xu and Xiaohui Fan), and the Mayall z-band Legacy Survey (MzLS; Prop. ID \#2016A-0453; PI: Arjun Dey). DECaLS, BASS and MzLS together include data obtained, respectively, at the Blanco telescope, Cerro Tololo Inter-American Observatory, NSF’s NOIRLab; the Bok telescope, Steward Observatory, University of Arizona; and the Mayall telescope, Kitt Peak National Observatory, NOIRLab. Pipeline processing and analyses of the data were supported by NOIRLab and the Lawrence Berkeley National Laboratory (LBNL). The Legacy Surveys project is honored to be permitted to conduct astronomical research on Iolkam Du’ag (Kitt Peak), a mountain with particular significance to the Tohono O’odham Nation.

NOIRLab is operated by the Association of Universities for Research in Astronomy (AURA) under a cooperative agreement with the National Science Foundation. LBNL is managed by the Regents of the University of California under contract to the U.S. Department of Energy.

This project used data obtained with the Dark Energy Camera (DECam), which was constructed by the Dark Energy Survey (DES) collaboration. Funding for the DES Projects has been provided by the U.S. Department of Energy, the U.S. National Science Foundation, the Ministry of Science and Education of Spain, the Science and Technology Facilities Council of the United Kingdom, the Higher Education Funding Council for England, the National Center for Supercomputing Applications at the University of Illinois at Urbana-Champaign, the Kavli Institute of Cosmological Physics at the University of Chicago, Center for Cosmology and Astro-Particle Physics at the Ohio State University, the Mitchell Institute for Fundamental Physics and Astronomy at Texas A\&M University, Financiadora de Estudos e Projetos, Fundacao Carlos Chagas Filho de Amparo, Financiadora de Estudos e Projetos, Fundacao Carlos Chagas Filho de Amparo a Pesquisa do Estado do Rio de Janeiro, Conselho Nacional de Desenvolvimento Cientifico e Tecnologico and the Ministerio da Ciencia, Tecnologia e Inovacao, the Deutsche Forschungsgemeinschaft and the Collaborating Institutions in the Dark Energy Survey. The Collaborating Institutions are Argonne National Laboratory, the University of California at Santa Cruz, the University of Cambridge, Centro de Investigaciones Energeticas, Medioambientales y Tecnologicas-Madrid, the University of Chicago, University College London, the DES-Brazil Consortium, the University of Edinburgh, the Eidgenossische Technische Hochschule (ETH) Zurich, Fermi National Accelerator Laboratory, the University of Illinois at Urbana-Champaign, the Institut de Ciencies de l’Espai (IEEC/CSIC), the Institut de Fisica d'Altes Energies, Lawrence Berkeley National Laboratory, the Ludwig Maximilians Universitat Munchen and the associated Excellence Cluster Universe, the University of Michigan, NSF’s NOIRLab, the University of Nottingham, the Ohio State University, the University of Pennsylvania, the University of Portsmouth, SLAC National Accelerator Laboratory, Stanford University, the University of Sussex, and Texas A\&M University.

BASS is a key project of the Telescope Access Program (TAP), which has been funded by the National Astronomical Observatories of China, the Chinese Academy of Sciences (the Strategic Priority Research Program “The Emergence of Cosmological Structures” Grant \# XDB09000000), and the Special Fund for Astronomy from the Ministry of Finance. The BASS is also supported by the External Cooperation Program of Chinese Academy of Sciences (Grant \# 114A11KYSB20160057), and Chinese National Natural Science Foundation (Grant \# 12120101003, \# 11433005).

The Legacy Survey team makes use of data products from the Near-Earth Object Wide-field Infrared Survey Explorer (NEOWISE), which is a project of the Jet Propulsion Laboratory/California Institute of Technology. NEOWISE is funded by the National Aeronautics and Space Administration.

The Legacy Surveys imaging of the DESI footprint is supported by the Director, Office of Science, Office of High Energy Physics of the U.S. Department of Energy under Contract No. DE-AC02-05CH1123, by the National Energy Research Scientific Computing Center, a DOE Office of Science User Facility under the same contract; and by the U.S. National Science Foundation, Division of Astronomical Sciences under Contract No. AST-0950945 to NOAO.

PEMP acknowledges the support from the Dutch Research Council (NWO) through the Veni grant VI.Veni.222.364.

KS acknowledges support from the Natural Sciences and Engineering Research Council of Canada (NSERC).

PK is supported by the BMBF project 05A20PC4 for D-MeerKAT.

JR acknowledges funding from University of La Laguna through the Margarita Salas Program from the Spanish Ministry of Universities ref. UNI/551/2021-May 26, and under the EU Next Generation.

LVM acknowledges financial support from the grant CEX2021-001131-S funded by MCIN/AEI/ 10.13039/501100011033, from the grant PID2021-123930OB-C21 funded by MCIN/AEI/ 10.13039/501100011033, by “ERDF A way of making Europe” and by the "European Union".

\section*{Data Availability}

The WALLABY source catalogue and associated data products (e.g. cubelets, moment maps, integrated spectra, radial surface density profiles) are available online through the CSIRO ASKAP Science Data Archive (CASDA) and the Canadian Astronomy Data Centre (CADC). All source and kinematic model data products are mirrored at both locations. Links to the data access services and the software tools used to produce the data products as well as documented instructions and example scripts for accessing the data are available from the WALLABY Data Portal.

This paper includes archived data obtained through the Australia Telescope Online Archive (http://atoa.atnf.csiro.au).
 



\bibliographystyle{mnras}
\bibliography{example} 




\appendix

\section{Kinematic Models}
\label{sec:bbarolo}

\begin{figure*}
     \centering
     \begin{subfigure}[t]{0.49\textwidth}
         \centering
         \includegraphics[width=\textwidth]{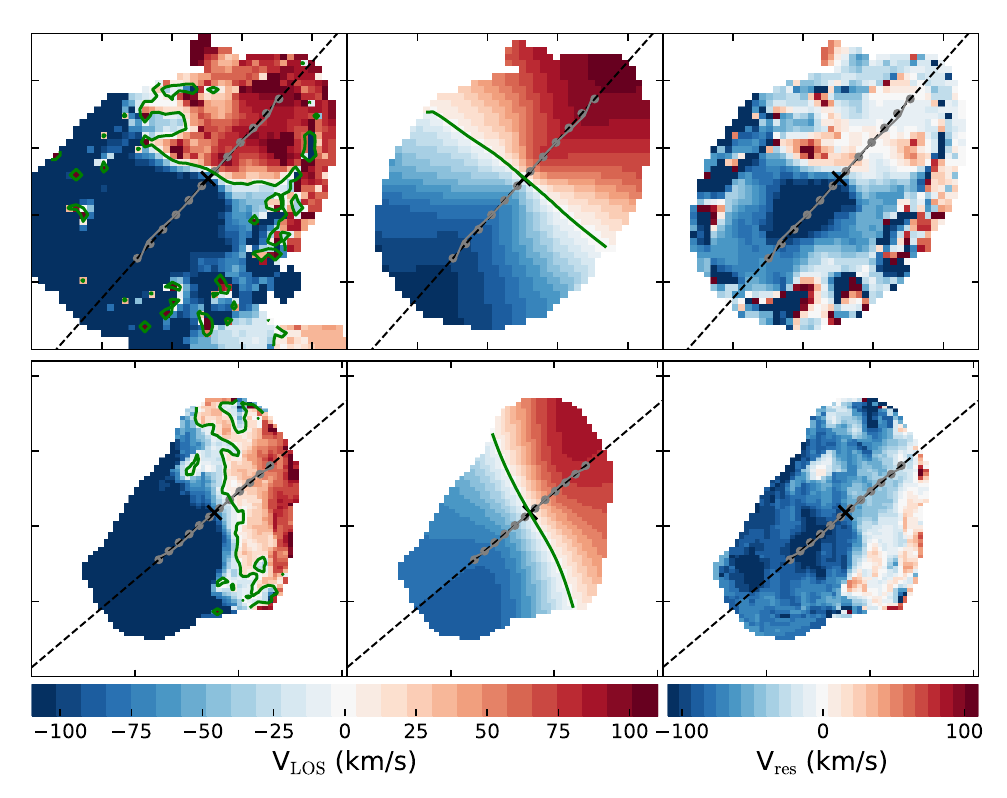}
         \caption{{\tt 3DBarolo} moment 1 maps. Top row: WALLABY, bottom row: ATCA. Left column: data, middle column: model, right column: residuals. The green contours show where the line of sight velocity is equal to zero.}
         \label{fig:m1blob}
     \end{subfigure}
     \hfill
     \begin{subfigure}[t]{0.49\textwidth}
         \centering
         \includegraphics[width=\textwidth]{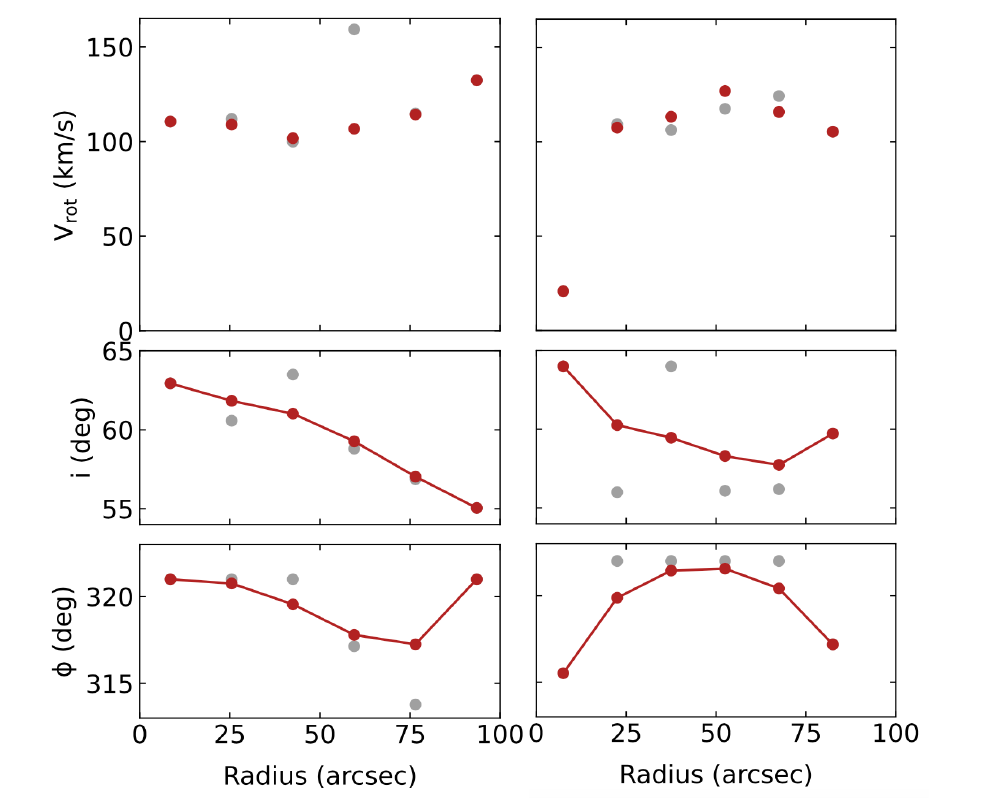}
         \caption{{\tt 3DBarolo} parameter fits. Left column: WALLABY, right column:ATCA. Top row: rotation velocity, middle row: inclination, bottom: position angle.}
         \label{fig:pltblob}
     \end{subfigure}
        \caption{{\tt 3DBarolo}models for ESO~437-G004 using WALLABY and ATCA data.}
        \label{fig:bbaroloblob}
\end{figure*}

\begin{figure*}
     \centering
     \begin{subfigure}[t]{0.49\textwidth}
         \centering
         \includegraphics[width=\textwidth]{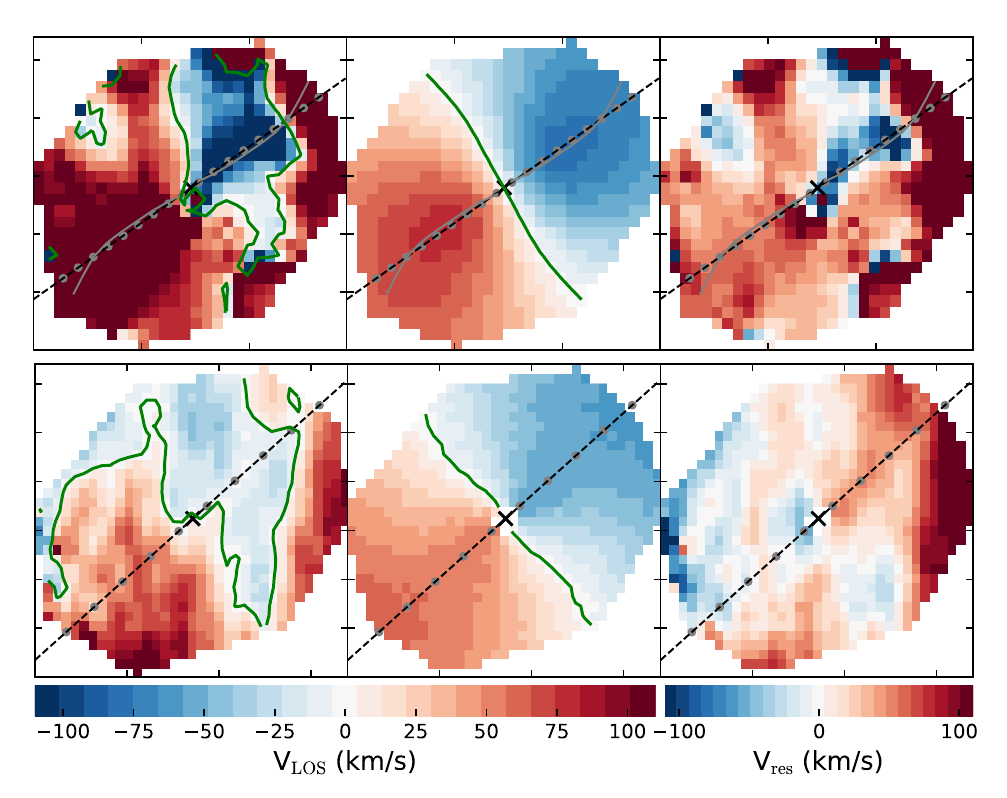}
         \caption{{\tt 3DBarolo} moment 1 maps. Top row: WALLABY, bottom row: ATCA. Left column: data, middle column: model, right column: residuals. The green contours show where the line of sight velocity is equal to zero.}
         \label{fig:m1spiral}
     \end{subfigure}
     \hfill
     \begin{subfigure}[t]{0.49\textwidth}
         \centering
         \includegraphics[width=\textwidth]{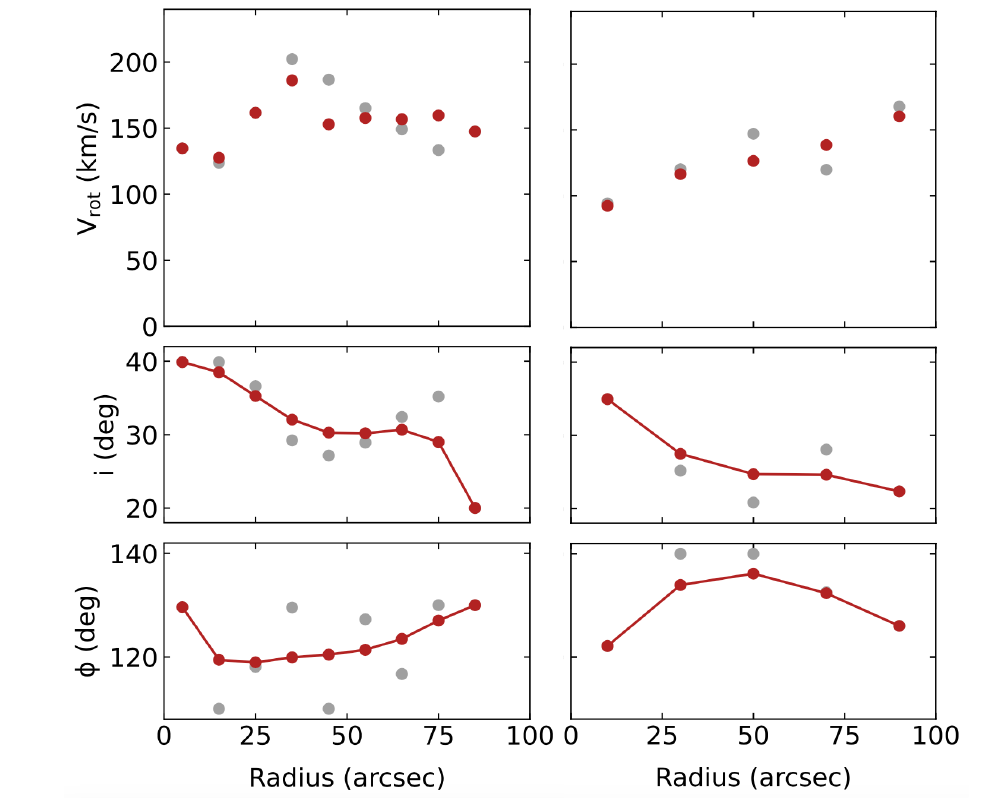}
         \caption{{\tt 3DBarolo} parameter fits. Left column: WALLABY, right column:ATCA. Top row: rotation velocity, middle row: inclination, bottom: position angle.}
         \label{fig:pltspiral}
     \end{subfigure}
        \caption{{\tt 3DBarolo} models for ESO~436-G046 using WALLABY and ATCA data.}
        \label{fig:bbarolospiral}
\end{figure*}

\begin{figure*}
    \centering
    \begin{subfigure}[]{0.49\textwidth}
        \includegraphics[width=\textwidth]{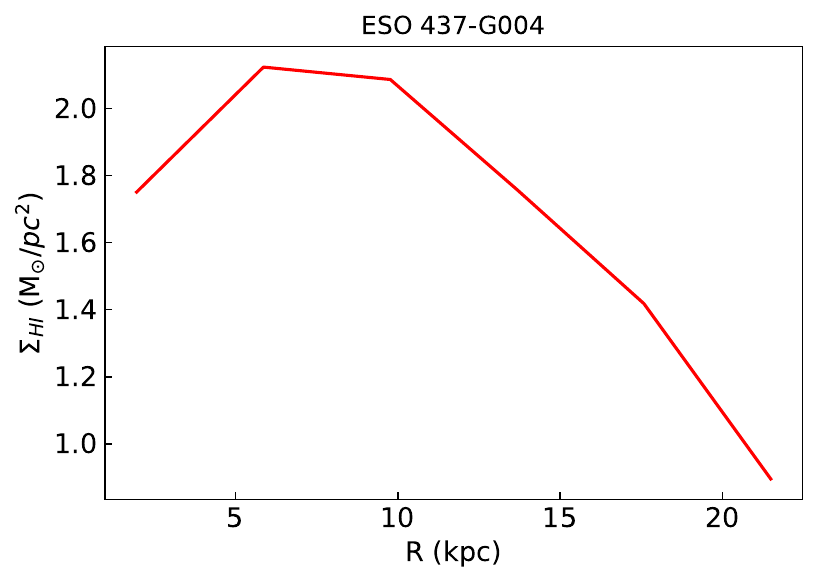}
        \caption{ }
        \label{fig:sigma_b}
    \end{subfigure}
    \begin{subfigure}[]{0.49\textwidth}
        \includegraphics[width=\textwidth]{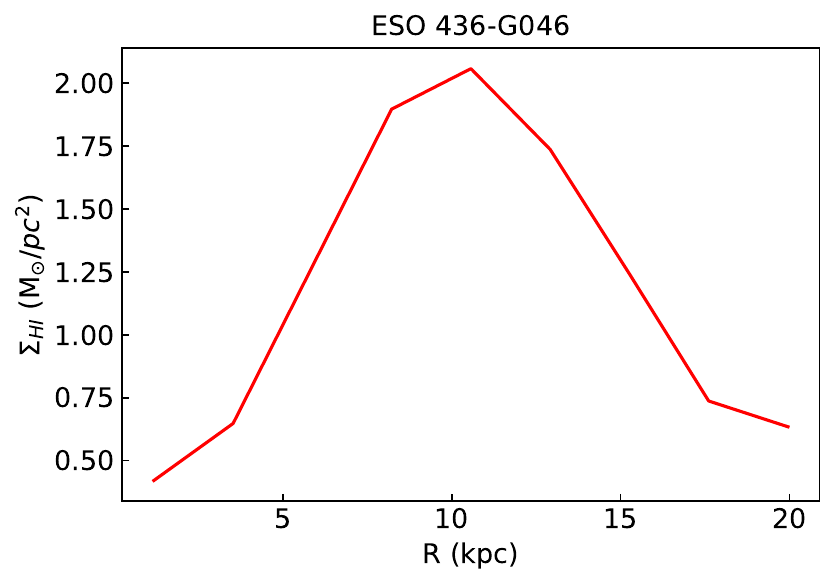}
        \caption{ }
        \label{fig:sigma_s}
    \end{subfigure}
    
    \caption{The H{\sc i} mass surface density profiles for ESO~437-G004 (left) and ESO~436-G046 (right) from the {\tt 3DBarolo} kinematic models. The WALLABY data was used.}
    \label{fig:density}
\end{figure*}

In this section we present the kinematic models of ESO~436-G046 and ESO~437-G004 created using {\tt 3DBarolo} \citep{DiTeodoro:15}. The software models the H{\sc i} gas as 3D tilted rings, fitting for rotation velocity, inclination and position angle, which can be seen in Figures \ref{fig:pltblob} and \ref{fig:pltspiral}. The red points in these figures are the final results, while the grey are the outputs of the first iteration.  Figure \ref{fig:bbaroloblob} and \ref{fig:sigma_b} shows the relevant outputs for ESO~437-G004 and \ref{fig:bbarolospiral} and \ref{fig:sigma_s} show those of ESO~436-G046.  The H{\sc i} mass surface density estimated by these kinematic models for the WALLABY data was used in the calculation of $S_{rps}$ in Figure~\ref{fig:Srps-r}. 

\begin{figure*}
    \centering
    \begin{subfigure}[t]{0.48\textwidth}
        \centering
        \includegraphics[width=0.99\textwidth]{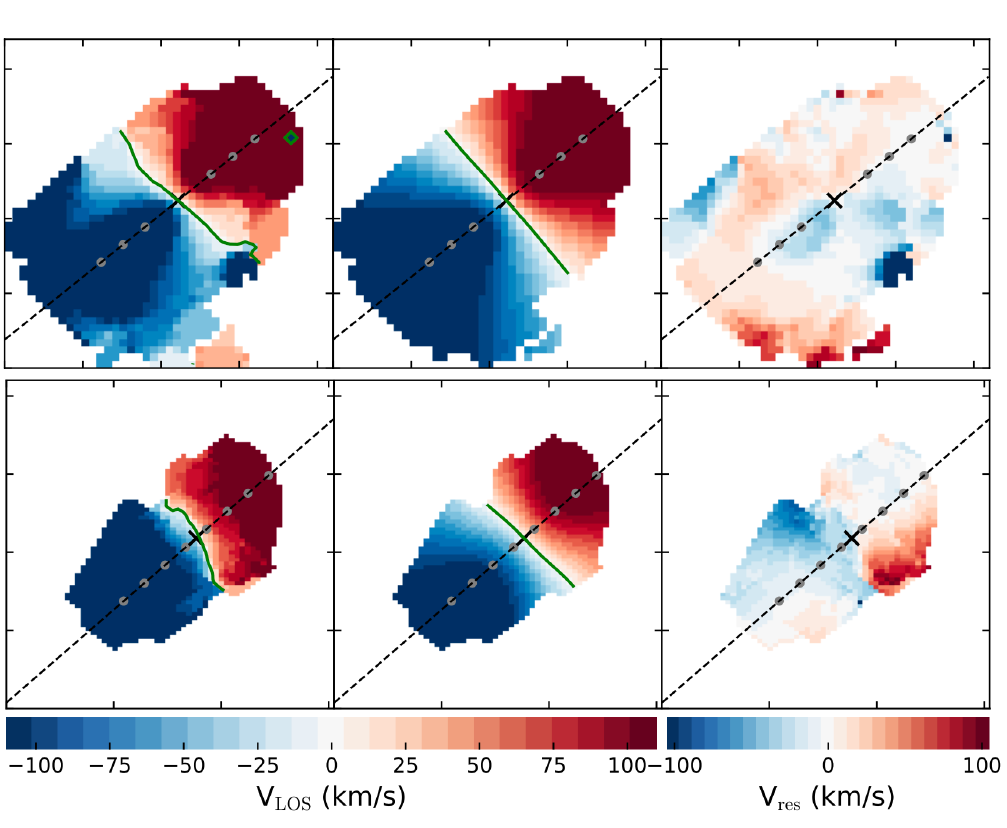}
        \caption{{\tt 3DBarolo} moment 1 maps. Top row: WALLABY, bottom row: ATCA. Left column: data, middle column: model, right column: residuals. The green contours show where the line of sight velocity is equal to zero.}
        \label{fig:pav_m1}
    \end{subfigure}
    \begin{subfigure}[t]{0.48\textwidth}
        \centering
        \includegraphics[width=0.99\textwidth]{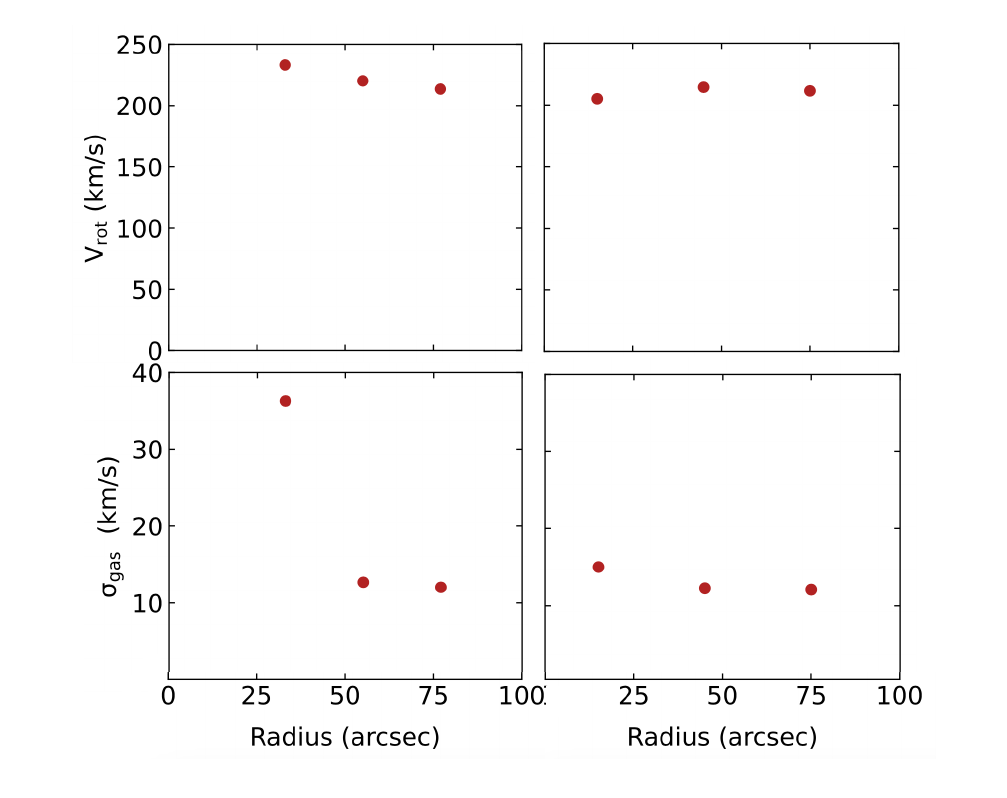}
        \caption{{\tt 3DBarolo} parameter fits. Left column: WALLABY, right column: ATCA. Top row: rotation velocity, bottom: velocity dispersion. The inclinations are set to 40$^{\circ}$ and the position angles are set to 310$^{\circ}$ and 320$^{\circ}$ for WALLABY and ATCA respectively.}
        \label{fig:pav_plt}
    \end{subfigure}
    
    \caption{Alternative {\tt 3DBarolo} models for ESO~437-G004 using WALLABY and ATCA data.}
    \label{fig:pavel}
\end{figure*}

It is worth noting here that the models are very sensitive to the initial estimates. To demonstrate this, alternative models for the WALLABY and ATCA data of ESO~437-G004 are shown in Figure \ref{fig:pavel}. In the original models, roughly two rings are fit per beam, as done in the WALLABY kinematic modelling pipeline \citep{Deg:22} and in previous WALLABY studies \citep{Elagali:19,Reynolds:19}. However, in the alternative models, the radial separation of the rings is increased to reduce the effects of oversampling. Additionally, these alternative models hold a fixed inclination and position angle, and fit for the rotation and dispersion velocity, whilst the original models fit for position angle, inclination and rotation velocity and hold the dispersion velocity fixed. Finally, the alternative models allow {\tt 3DBarolo} to find its own mask in the already masked cube. These changes to initial conditions result in rotation velocities of $\sim220$ km s$^{-1}$ and $\sim210$ km s$^{-1}$ for the WALLABY and ATCA data respectively. Not only is this significantly higher than the rotation velocities of the original model, but also is significantly higher than the WKAPP value. If we use the results of the alternative models, then the extended gas of ESO~437-G004 would be far less susceptible to ram pressure stripping. On average, the residuals of the alternate models are a factor of $\sim 3$ lower than those of the original models. The impact of changing the initial conditions is not unexpected, as \cite{Kamphuis:15} found that kinematic modelling for observations with fewer than eight beams will likely lead to large degeneracies in parameters and loose constraints. Hence, future deeper and higher resolution observations are likely to improve our constraints on the kinematic modelling.

\section{Stellar Mass and Specific Star Formation Rate}

The values used in the calculations of the stellar mass and specific star formation rate in Section \ref{sec:mstar} are shown in Table \ref{tab:mstarsfr}. The references for this data and the equations used are presented in Section \ref{sec:mstar}.

\begin{table*}[h!]
    \centering
    \caption{ The quantities used in the calculations of stellar mass and specific star formation rate in Section \ref{sec:mstar}. The rows from top to bottom are:  \textit{GALEX} NUV flux density, W1 \textit{WISE} magnitude, W2 \textit{WISE} magnitude, W3 \textit{WISE} magnitude, and W4 \textit{WISE} magnitude.} 
    \label{tab:mstarsfr}
    \begin{tabular}{lcccccccc}
        \hline
         Parameter & ESO & ESO & ESO-LV & PGC  & PGC & ESO & ESO & ESO \\
          & 436-G046 & 436-IG042 & 4360421 & 031288 & 031270 & 436-G044 & 436-G045 & 437-G004 \\
         \hline
         $F_{\rm NUV}$ ($\mu$Jy) & $8686\pm45$ & $771\pm20$ & - & $7\pm3$ & $224\pm7$ & $95\pm6$ & $37\pm4$ & $947\pm17$ \\
         $m_{\rm W1}$ (mag) & $11.23\pm0.02$ & $11.11\pm0.02$ & $13.40\pm0.02$ & $11.30\pm0.02$ & $12.81\pm0.02$ & $10.33\pm0.02$ & $11.25\pm0.02$ & $11.42\pm0.02$ \\
         $m_{\rm W2}$ (mag) & $11.26\pm0.02$ & $10.74\pm0.02$ & $13.42\pm0.03$& $11.37\pm0.02$ & $12.78\pm0.03$ & $10.38\pm0.02$ & $11.34\pm0.02$ & $11.38\pm0.02$  \\
         $m_{\rm W3}$ (mag) & - & - & - & $10.68\pm0.09$ & - & - & $10.78\pm0.09$ & - \\
         $m_{\rm W4}$ (mag) & $6.02\pm0.04$ & $3.31\pm0.03$ & $8.56\pm0.34$ & - & $7.02\pm0.08$ & $8.31\pm0.26$ & - & $5.79\pm0.05$ \\
         \hline
    \end{tabular}

\end{table*}


\bsp	
\label{lastpage}
\end{document}